%% file: naturalness.tex
\documentclass[onecolumn, 11pt, showpacs, nofootinbib, amsmath, amssymb, aps, prd, longbibliography, floatfix]{revtex4-1}
\pdfoutput=1

\usepackage{amsmath}
\usepackage{color}
\usepackage{epsfig}
\usepackage{graphics}
\usepackage{graphicx}
\usepackage{hyperref}
\usepackage{silence}
\usepackage{slashed}
\usepackage[caption=false]{subfig}

\input{tikzdefs.tex}
\newif\iftikz
\tikztrue

\ErrorFilter{amsmath}{Erroneous nesting of equation structures}

\newcommand{\eq}[1]{Eq.~(\ref{#1})}
\newcommand{\eqs}[1]{Eqs.~(\ref{#1})}
\newcommand{\fig}[1]{Fig.~\ref{#1}}
\newcommand{\figs}[1]{Figs.~\ref{#1}}
\newcommand{\tab}[1]{Table~\ref{#1}}
\newcommand{\tabs}[1]{Tables~\ref{#1}}
\renewcommand{\sec}[1]{Sec.~\ref{#1}}
\newcommand{\citeref}[1]{Ref.~\cite{#1}}
\newcommand{\citerefs}[1]{Refs.~\cite{#1}}

\begin{document}

\title{Naturalness made easy: \texorpdfstring{\\}{} two-loop naturalness bounds on minimal SM extensions}

\author{Jackson D. Clarke, Peter Cox}
\affiliation{\small ARC Centre of Excellence for Particle Physics at the Terascale, \\ 
School of Physics, University of Melbourne, 3010, Australia.}

\begin{abstract}

The main result of this paper is a collection of conservative naturalness bounds on minimal 
extensions of the Standard Model by (vector-like) fermionic or scalar gauge multiplets.
Within, we advocate for an intuitive and physical concept of naturalness
built upon the renormalisation group equations.
In the effective field theory 
of the Standard Model plus a gauge multiplet with mass $M$,
the low scale Higgs mass parameter
is a calculable function of $\overline{\rm MS}$ input parameters
defined at some high scale $\Lambda_h > M$.
If the Higgs mass is very sensitive to these input parameters,
then this signifies a naturalness problem.
To sensibly capture the sensitivity, 
it is shown how a sensitivity measure can be rigorously derived
as a Bayesian model comparison, 
which reduces in a relevant limit to a Barbieri--Giudice-like fine-tuning measure.
This measure is fully generalisable to any perturbative EFT.
The interesting results of our two-loop renormalisation group study are as follows:
for $\Lambda_h=\Lambda_{Pl}$ we find ``$10\%$ fine-tuning'' bounds on the masses
of various gauge multiplets of $M<\mathcal{O}(1$--$10)$~TeV, with
bounds on fermionic gauge multiplets significantly weaker than for scalars;
these bounds remain finite in the limit $\Lambda_h\to M^+$,
weakening to $M<\mathcal{O}(10$--$100)$~TeV;
and bounds on coloured multiplets are no more severe than for electroweak multiplets,
since they only directly correct the Higgs mass at three-loop.

\end{abstract}


\maketitle

\section{Introduction}

The Standard Model (SM) appears to represent a very good effective field theory (EFT) for energies at least $\lesssim$~TeV.
Still, it has several well known theoretical and phenomenological shortcomings.
Many of these can be addressed
with minimal extensions of the SM by
heavy fermionic and/or scalar gauge multiplets (GMs).
However, the Higgs mass parameter $\mu^2(m_Z)\approx -(88\text{ GeV})^2$,
appearing in the SM potential $\mu^2 H^\dagger H + \lambda (H^\dagger H)^2$,
is sensitive to such heavy new physics;
GMs couple (at the very least) at loop level to
the SM Higgs, thereby inducing corrections to the Higgs mass
and potentially introducing a naturalness problem.

The subject of naturalness in the modern literature is rife with various (and often conflicting) definitions.
Let us therefore, at the outset, state the definition used in this paper:
{\it a parameter in a quantum field theory is ``natural'' if 
its measured value at low scale
is (sufficiently) insensitive to details of the physics at high scale}.
Plainly, then, to examine naturalness of the Higgs mass parameter we require:
(1) a description of the low scale physics; 
(2) a description of the high scale physics;
(3) a map which relates them;
and (4) a measure which quantifies sensitivity of $\mu^2(m_Z)$ to the high scale physics.

In this paper we confront the question,
{\it at what mass does a heavy GM introduce a 
physical Higgs naturalness problem?}
Vector-like fermionic and scalar GMs of various charges are studied.
We advocate a renormalisation group (RG) approach to naturalness.
The description of the low (high) scale physics is provided by the 
$\overline{{\rm MS}}$ Lagrangian parameters of the SM (SM+GM) EFT 
defined at the scale $m_Z$ ($\Lambda_h$),
and the map which relates them is the set of RG equations (RGEs).
We employ a sensitivity measure which can be interpreted 
as a Bayesian model comparison. Several Bayesian approaches to naturalness 
have previously been considered in the literature~\cite{Strumia:1999fr,Allanach:2006jc,Allanach:2007qk,Athron:2007ry,Cabrera:2008tj,
AbdusSalam:2009qd,Cabrera:2009dm,Ghilencea:2012gz,Fichet2012sn,Kim:2013uxa,
Fowlie:2014xha,Fowlie:2014faa,Fowlie:2016jlx}. Inspired by the approach of \citeref{Fichet2012sn}, we propose a 
particular model comparison which captures the ``naturalness price'' paid
for promoting the Higgs mass parameter from a purely phenomenological input
parameter at low scale to a high scale input parameter of the model. We
show that this sensitivity measure then reduces in a well-motivated
limit to a Barbieri--Giudice-like \cite{Barbieri1987fn,Ellis1986yg} fine-tuning measure.
Quantifying and bounding this sensitivity results in naturalness bounds
on the masses of the various GMs.

There exist many phenomenologically motivated extensions to the SM which involve just a single GM. 
Such models can address a wide variety of shortcomings of the SM including 
neutrino mass~\cite{Magg:1980ut,*Lazarides:1980nt,*Schechter:1980gr,*Cheng:1980qt,*Mohapatra:1980yp,Foot:1988aq,Ma:2006km}, 
dark matter~\cite{Barbieri:2006dq,Cirelli:2005uq}, baryogenesis~\cite{Fromme:2006cm} 
and the strong CP problem~\cite{Kim:1979if,*Shifman:1979if,Zhitnitsky:1980tq,*Dine:1981rt}.
The naturalness bounds we derive can be applied to these models and many others besides. 
Our results can also be used to provide a qualitative, conservative bound even in extended models. Moreover, the framework 
we present is completely general and could in principle be applied to any model.

The paper is organised as follows.
In \sec{SecNaturalness} we describe an intuitive and physical 
concept of naturalness built upon the RGEs,
and the sensitivity measure as a Bayesian model comparison.
In \sec{SecMethod} we describe how this concept is applied to the SM+GM EFTs.
Our main result is a list of naturalness bounds presented in \sec{SecResults}.
These results are discussed in \sec{SecDiscussion},
and we conclude in \sec{SecConclusion}.

\section{Physical Naturalness \label{SecNaturalness}}

In this section we describe a
physical way to understand the Higgs naturalness problem,
especially pertinent in the context of bottom-up extensions of the SM.
To frame the discussion, let us appeal to an illuminating toy model.

\subsection{Toy model \label{SecExample}}

Consider a perturbative EFT consisting of the 
SM plus a heavy particle of mass $M$
(whose mass is obtained independently of electroweak symmetry breaking).
The $\mu^2$ RGE valid at a renormalisation scale $\mu_R > M$
takes the form
\begin{align}
 \frac{d}{d\log\mu_R}\mu^2(\mu_R) = C_1(\mu_R) \mu^2(\mu_R) + C_2(\mu_R) M^2(\mu_R)\ , \label{EqRGE}
\end{align}
where $C_1(\mu_R) \simeq 6 y_t(\mu_R)/(4\pi)^2$, 
with $y_t$ the top quark Yukawa coupling.
The quantity $C_2(\mu_R)$ might be comprised of SM and/or beyond-SM couplings.
The RGEs allow $\mu^2$ (and other low scale parameters)
to be extrapolated to a high scale $\Lambda_h$,
at most up to the scale at which the EFT is no longer valid.
At $\Lambda_h$, these renormalised parameters 
can be interpreted as ``input parameters''
which might be derived from even higher scale physics.
The input parameters are, by construction, connected with 
the low energy parameters via the RGEs.
If the low energy parameters are very sensitive to these input parameters,
then this signifies a naturalness problem.

Let us now, under this paradigm,
try to understand when a heavy particle introduces a Higgs naturalness problem.
One can fully solve \eq{EqRGE} in the limit where 
$C_1$, $C_2$, and $M^2$ have no scale dependence.
Including a possible threshold correction, $\mu^2_+(M)=\mu^2_-(M)-C_T M^2$, 
when the SM EFT parameters ($-$) are matched onto the full EFT parameters ($+$) at the threshold $M$,
and in the limit $C_1 \log(\Lambda_h/m_Z) \ll 1$,
\begin{align}
 \mu^2(m_Z) \simeq  \  
 \mu^2(\Lambda_h) - \Theta(\Lambda_h - M) \left[ 
  C_2 M^2 \log \left( \frac{\Lambda_h}{M} \right)
  - C_T M^2
  \right], \label{Eqmu2mZ}
\end{align}
where $\Theta$ is the step function.
It is now easy to see when a naturalness problem arises.
If either of $C_2M^2$ or $C_T M^2$ is $\gg \mu^2(m_Z)$, 
then the input parameter $\mu^2(\Lambda_h)$ must be finely tuned 
against a very large contribution
in order to realise the observed Higgs mass.
A small change in $\mu^2(\Lambda_h)$ ruins this cancellation,
and thus the Higgs mass is unnatural, 
i.e. it is sensitive to details of the high scale physics.
Note that the $C_2M^2$ piece captures a steep $\mu^2(\mu_R)$ RG trajectory,
whereby only a very particular input $\mu^2(\Lambda_h)$
will lead to the observed low scale $\mu^2(m_Z)$;
a small change in this value leads to significant over- or under-shooting.

In this picture, the Higgs naturalness problem is cast in terms of a 
potential sensitivity between {\it measurable} parameters, 
connected by fully calculable (in a perturbative theory)
RG trajectories and matching conditions.
The picture shares aspects with other discussions 
which have appeared in the literature, e.g. \citerefs{Wetterich:1983bi,Bardeen:1995kv,Aoki:2012xs,Farina:2013mla,Heikinheimo:2013fta,
Bian:2013xra,deGouvea:2014xba,Kobakhidze:2014afa,Fabbrichesi:2015zna,Fujikawa:2016hvy}.
In particular,
the large cancellation between the unmeasurable bare mass
and the cutoff regulator contribution is considered
an unphysical artifact of the regularisation and renormalisation procedure for the EFT.
The naturalness bounds we will derive within this picture
ostensibly most resemble those in \citeref{Farina:2013mla},
however they differ in a few key respects which we highlight here.
First, we include the full effect of RGE running and matching, rather than simply 
bounding the finite loop corrections to $\mu^2$.
Second, since the parameters which appear in the RGEs and matching conditions
are treated equally in our framework,
the sensitivity arising from all parameters (and not just the gauge terms) 
are automatically included.
Third, the appearance of apparently natural ``throats,''
or miraculous cancellations among these corrections,
are argued to be generically erased by 
the RG evolution and matching conditions.
Last, we include (for the first time as far as we are aware) 
the leading three-loop correction
for vector-like fermions. The inclusion of these effects leads, in most cases, 
to weaker naturalness bounds on the masses of heavy GMs than those obtained in~\citeref{Farina:2013mla}.

\subsection{Sensitivity measure}

To actually quantify the sensitivity to high scale physics 
can seem somewhat arbitrary and subjective.
There are many approaches in the literature.
It is important to appreciate that 
assumptions about the unknown high scale physics
necessarily enter into all of these approaches.
However these assumptions are seldom explicitly stated,
and the fine-tuning or sensitivity measures they motivate
are often only written down intuitively.
These shortcomings are overcome in the 
Bayesian approach to naturalness which we utilise here.
In this approach, assumptions about the high scale physics
can be explicitly and clearly stated, 
appearing as a complete set
of prior densities for the high scale parameters.
The requirement that our result be insensitive
to units or parameter rescalings defines a set of agnostic 
priors, and the resulting naturalness bounds
turn out to be rather insensitive to departures from these priors
(this will be discussed in more detail in \sec{SecDiscussion}).
Further, instead of being written down directly,
the sensitivity measure is rigorously derived from an underlying framework,
having a well-defined interpretation as a Bayesian model comparison.
The measure also reduces (in the cases we consider) to a
Barbieri--Giudice-like \cite{Barbieri1987fn,Ellis1986yg} fine-tuning measure,
which can be easily understood intuitively.
We provide below a short description; 
more details can be found in Appendix~\ref{AppBayes}.

Assuming a flat prior belief in the 
high scale input parameters $\mathcal{I}=(\mathcal{I}_1,\dots,\mathcal{I}_n)$,
and a perfectly measured set of $m\le n$ independent observables $\mathcal{O}=(\mathcal{O}_1,\dots,\mathcal{O}_m)$, 
the Bayesian evidence $B$ for a model $\mathcal{M}$ 
is a function of the unconstrained input parameters 
$\mathcal{I}'=(\mathcal{I}_{m+1},\dots,\mathcal{I}_n)$:
\begin{align}
 \left.
 B(\mathcal{M};\mathcal{I}') \propto
  \frac{1}{\sqrt{\left| J J^T \right|}}
  \ \right|_{
  ^{\mathcal{O}_{ex}}
  _{\mathcal{I}'}
  }  \ ,
\end{align}
where $J$ is the $m\times n$ matrix 
defined by $J_{ij} = \partial\mathcal{O}_i / \partial\mathcal{I}_j$~\cite{Fichet2012sn}.
Let us take, for model $\mathcal{M}$, $\mathcal{I}_1=\log\mu^2(\Lambda_h)$ and $\mathcal{O}_1=\log\mu^2(m_Z)$.
The logarithms here ensure that our result is independent with respect to units or parameter rescalings
(absolute values are implied for the argument of any $\log$ and dimensionful parameters can be normalised by any unit).
Our Higgs mass sensitivity measure arises from a particular Bayesian model comparison:
we compare to a model $\mathcal{M}_0$ 
in which we instead take 
$\mathcal{I}_1=\mathcal{O}_1=\log\mu^2(m_Z)$, i.e. the Higgs mass parameter is considered as an input parameter at scale $m_Z$.
The sensitivity measure can then be written as a function of the unconstrained parameters,
\begin{align}
 \Delta(\mathcal{M};\mathcal{I}') =
 \frac{B(\mathcal{M}_0;\mathcal{I}')}{B(\mathcal{M};\mathcal{I}')}  \ .
 \label{EqFTuning0}
\end{align}
This measure captures the ``naturalness price'' 
paid for promoting the Higgs mass parameter
to a high scale input parameter of the model
as opposed to a purely phenomenological input parameter at low scale.
In our context, a large value of $\Delta$ essentially tells us that,
given a flat prior density in $\log\mu^2(\Lambda_h)$,
the observed value $\mu^2(m_Z)$ is unlikely
[specifically with respect to a flat probability density in $\log \mu^2(m_Z)$],
i.e. $\mu^2(m_Z)$ is sensitive to the realised input parameters.
In the special case that the low scale observables,
except for possibly $\mu^2(m_Z)$,
are approximately insensitive to the unconstrained inputs,
$B(\mathcal{M}_0;\mathcal{I}')$ becomes independent of $\mathcal{I}'$ and
\eq{EqFTuning0} reduces to
\begin{align}
 \Delta(\mathcal{M};\mathcal{I}') \simeq \left.
 \sqrt{
   \left( \frac{\partial\log\mu^2(m_Z)}{\partial\log\mu^2(\Lambda_h)} \right)^2
   + \sum\limits_{j\ge m+1} \left( \frac{\partial\log\mu^2(m_Z)}{\partial\mathcal{I}_j} \right)^2
   }
  \right|_{
  ^{\mathcal{O}_{ex}}
  _{\mathcal{I}'}
  } \ .
 \label{EqFTuning2}
\end{align}
In the absence of unconstrained inputs ($n=m$) the summation disappears and the equality is exact.
This is clearly reminiscent of the Barbieri--Giudice fine-tuning measure.
A value of $\Delta=10$ can now be interpreted as
the onset of strong Bayesian evidence 
(for $\mathcal{M}_0$ over $\mathcal{M}$) 
on the Jeffreys scale \cite{Jeffreys1939xee},
or a 10\% fine-tuning from the Barbieri--Giudice perspective.

Notice here that we have a sensitivity measure which depends on unconstrained inputs.
It might be that we want to ``project out'' some of these nuisance parameters.
In this paper we will minimise over them,
which picks out a conservative best case naturalness scenario in the model.
Our SM+GM models $\mathcal{M}$ are defined 
by $\overline{\rm MS}$ inputs at the high scale $\Lambda_h$,
with the renormalised mass parameter $M^k(\Lambda_h)\subset \mathcal{I}'$ 
[and $k=1$ (2) in the fermionic (scalar) case].
We minimise over all unconstrained parameters apart from $M^k(\Lambda_h)$ 
to obtain a sensitivity measure which depends only on $M$ and $\Lambda_h$:
\begin{align}
 \Delta(M,\Lambda_h) =&\ \min\limits_{\mathcal{I}'\setminus \{M^k(\Lambda_h)\}} 
 \bigg[
 \Delta(\mathcal{M};\mathcal{I}')
 \bigg] \ .
 \label{EqFTuning1}
\end{align}
In practice we minimise over \eq{EqFTuning2}, 
which is now valid under the looser criterion that
the low scale observables,
except for possibly $\mu^2(m_Z)$,
are approximately insensitive to the unconstrained inputs
in the vicinity of the minimum.

This all may sound rather abstract.
Let us now check that, in the relevant cases, 
the sensitivity measure \eq{EqFTuning1} captures 
the sensitivity we expect in our toy model when $C_2M^2,C_TM^2\gg \mu^2(m_Z)$.

\subsection{Fermion-like case \label{Sec2Fermion}}

In the minimal fermionic SM+GM there are no new dimensionless parameters; 
$C_2$ is fully constrained by experiment so that
$\mathcal{O}_i=\left\{ \log\mu^2(m_Z), \log C_1, \log C_2 \right\}$
and $\mathcal{I}_j=\left\{ \log\mu^2(\Lambda_h), \log C_1, \log C_2, \log M \right\}$.
It is easy to show directly from \eq{EqFTuning0} that, even allowing for possible $C_{1,2}$ correlation
$\partial\log C_1/\partial\log C_2 \ne 0$,
\begin{align}
  \Delta(M,\Lambda_h)
= \sqrt{ \left(\frac{\partial\log\mu^2(m_Z)}{\partial\log\mu^2(\Lambda_h)}\right)^2 +
         \left(\frac{\partial\log\mu^2(m_Z)}{\partial\log M}\right)^2 }.
\label{EqDeltaToyModelF}
\end{align}
This is just a Barbieri--Giudice-like fine-tuning measure
comparing percentage changes in the low scale Higgs mass parameter 
to those in the input parameters.
In the limit $C_1 \log(\Lambda_h/m_Z) \ll 1$ and taking $\Lambda_h>M$,
we see that $\Delta(\Lambda_h)$ is made up of two pieces:
\begin{align}
 \left| \frac{\partial\log\mu^2(m_Z)}{\partial\log\mu^2(\Lambda_h)} \right| = &\ 
  \left| 1+\frac{C_2M^2}{\mu^2(m_Z)}\log\left(\frac{\Lambda_h}{M}\right) - \frac{C_TM^2}{\mu^2(m_Z)} \right| , \nonumber \\
  k \left| \frac{\partial\log\mu^2(m_Z)}{\partial\log M^k} \right| = &\ 
  \left| \frac{C_2M^2}{\mu^2(m_Z)}\left[2 \log\left(\frac{\Lambda_h}{M}\right)-1 \right] - 2\frac{C_TM^2}{\mu^2(m_Z)} \right| .
  \label{EqDeltaToyModelF2}
\end{align}
The ``1'' piece is the SM contribution,
the $\log(\Lambda_h/M)$ pieces
reflect sensitivity to the RG trajectory (with slope $C_2M^2$),
the $C_TM^2$ piece is due to the finite threshold correction,
and the log-independent $C_2M^2$ piece arises because 
a variation in $\log M$ results in a shift in the matching scale,
which reintroduces a small amount of RG evolution.
Clearly $\Delta\gg 1$ if $C_2M^2$ or $C_TM^2$ is $\gg \mu^2(m_Z)$, as expected.
This even holds in the limit where the high scale approaches the heavy particle mass, $\Lambda_h\to M^+$.

In the fermionic SM+GM EFT at two-loop order with one-loop matching, we have
\begin{align}
 C_2\sim \frac{g^4}{(4\pi)^4}, && C_T=0,
\end{align}
where $g$ is a placeholder for a gauge coupling(s).
It is interesting to note that in the limit $\Lambda_h\to M^+$,
\eq{EqDeltaToyModelF} just becomes
\begin{equation}
\Delta(M^+)=\sqrt{1+\left(\frac{C_2(M) M^2}{\mu^2(m_Z)}\right)^2}\ .
\end{equation}
If we bound this sensitivity measure by $\Delta_{max}$,
this is almost equivalent to simply bounding the contribution to the $\mu^2(\mu_R)$ RGE in \eq{EqRGE}
at the scale $\mu_R=M$, i.e.
$C_2(M) M^2 \lesssim \Delta_{max} \mu^2(m_Z)$. 
This is not an uncommon practice as a zeroth-order naturalness bound for $M$.

There is one case we wish to comment on here:
the special case where $C_2(M)$ happens to vanish,
somewhat reminiscent of the Veltman condition \cite{Veltman:1980mj}.
In this case there is plainly no naturalness bound on $M$ from the $\Delta(M^+)$ measure
we have written above, no matter the size of $M$.
So it appears that there is a fine-tuning which is not captured by our framework in this limit.
Is this indeed the case?
One can show that extending this toy model to include 
RG evolution of $C_2$ is not enough to reintroduce the naturalness bound
(we will see this in our numerical analysis).
Instead, it turns out that this apparent ``Veltman throat'' 
is only a limitation of the order to which we are working.
In the fermionic SM+GM EFT with {\it two}-loop matching, 
$C_T$ becomes a non-zero function of the gauge couplings.
In general $C_T(M)\ne 0$ when $C_2(M)=0$,
and thus a sensitivity proportional to $M^2$ and powers of gauge couplings 
is recaptured at this special value of $M$. 
In any case, we do not attribute much physical significance to this special case; 
in the full model, even if $C_2(M)=0$,
RG effects reinstate $C_2(\mu_R)\ne 0$ at $\mu_R>M$,
and the sensitivity of $\mu^2(m_Z)$ to $M^2$ is {\it rapidly} 
recaptured in the realistic case with $\Lambda_h>M$.

\subsection{Scalar-like case}

Let us first consider the SM plus scalar GM case with only one portal quartic
$\lambda_{H1} (H^\dagger H)(\Phi^\dagger\Phi)$
and one self quartic
$\lambda_\Phi (\Phi^\dagger\Phi)(\Phi^\dagger\Phi)$.
This occurs whenever the scalar is an $SU(2)$ singlet.
At two-loop with one-loop matching, we have
\begin{align}
 C_2 = C_2^{SM} + 2 Q_3
  \frac{\lambda_{H1}}{(4\pi)^2} + \dots, &&
 C_T = Q_3
  \frac{\lambda_{H1}}{(4\pi)^2} ,
\end{align}
where $C_2^{SM}\sim g^4/(4\pi)^4$, and $(Q_1, Q_2, Q_3)$ are the $(U(1)_Y, SU(2), SU(3))$ charges of the GM.
Assuming for simplicity no RG evolution of these parameters, we have
$\mathcal{O}_i=\{ \log\mu^2(m_Z),$ $\log C_1,$ $\log C_2^{SM} \}$
and $\mathcal{I}_j=\left\{ \log\mu^2(\Lambda_h), \log C_1, \log C_2^{SM}, \log\lambda_{H1}, \log\lambda_\Phi, \log M^2 \right\}$.
The sensitivity measure,
assuming $C_1$ and $C_2^{SM}$ are insensitive to changes in $\lambda_{H1}$ and $\lambda_\Phi$,
is given by
\begin{align}
  \Delta(M,\Lambda_h) = \min\limits_{\lambda_{H1}} 
 \left[\left.
 \sqrt{ \left(\frac{\partial\log\mu^2(m_Z)}{\partial\log\mu^2(\Lambda_h)}\right)^2 +
         \left(\frac{\partial\log\mu^2(m_Z)}{\partial\log M^2}\right)^2  +
         \left(\frac{\partial\log\mu^2(m_Z)}{\partial\log \lambda_{H1}}\right)^2 }\right|_{\lambda_{H1}} 
         \right] , \label{EqSensitivityScalarLike1}
\end{align}
where we have ignored the subdominant $\partial/\partial\log\lambda_\Phi$ term for clarity.
Before minimisation, the contribution of the first two terms under the square root are exactly those in \eq{EqDeltaToyModelF2}.
Note that the $C_2 M^2$ contributions
are removed if $\lambda_{H1}$ takes the fortuitous value $-(4\pi)^2 C_2^{SM} / (2Q_3)$,
however the $C_TM^2$ contributions remain.
Conversely, $\lambda_{H1}=0$ removes the threshold correction contributions,
leaving non-vanishing $C_2 M^2$ contributions.
Thus it seems that, even with the extra freedom granted by $\lambda_{H1}$, 
one cannot remove the naturalness problem.
Indeed, the minimisation over $\lambda_{H1}$ can be performed analytically in this toy model. 
The result is rather lengthy, and we do not reproduce it here.
It is anyway not terribly illuminating,
since this toy model is too strong of an oversimplification to reflect 
the full scalar SM+GM case when $\Lambda_h\gg M$;
pure gauge contributions in the $\lambda_{H1}(\mu_R)$ RGE
which destabilise any fortuitous cancellation for $C_2(\mu_R)=0$ must be taken into account.
Nevertheless, the toy model result can serve as an argument for the existence
of a finite naturalness bound even in the limit $\Lambda_h\to M^+$, where we obtain
\begin{align}
 \Delta(M^+) = \sqrt{
  \frac{1}{12}\left[
  10 + 4 \frac{C_2^{SM} M^2}{\mu^2(m_Z)} + \left( \frac{C_2^{SM} M^2}{\mu^2(m_Z)}\right)^2
  \right]
 },
\end{align}
if $C_1 \log(\Lambda_h/m_Z) \ll 1$.
Again, clearly $\Delta \gg 1$ when $C_2^{SM} M^2\gg \mu^2(m_Z)$, as expected.
If we bound this sensitivity measure by $\Delta(M^+)<\Delta_{max}$,
this is approximately equivalent to $C_2^{SM}(M) M^2/\sqrt{12} \lesssim \Delta_{max} \mu^2(m_Z)$.

The scalar $SU(2)$ doublet or triplet SM+GM EFT case with two portal quartic couplings is more delicate.
At two-loop with one-loop matching we have
\begin{align}
 C_2 = C_2^{SM} + 2 Q_3 Q_2 
  \frac{\lambda_{H1}}{(4\pi)^2} - 24 Q_3 \frac{\lambda_{H2}^2}{(4\pi)^4} + \dots, &&
 C_T = Q_3 Q_2 
  \frac{\lambda_{H1}}{(4\pi)^2} , \label{EqC2SScalar2}
\end{align}
where $\lambda_{H1,2}$ will be defined in \sec{SecMethodScalar}.
There is now enough freedom for a minimisation analogous to \eq{EqSensitivityScalarLike1} 
to select $\lambda_{H1,2}$ such that $C_2=0$ and $C_T=0$ simultaneously,
removing the sensitivity of $\mu^2(m_Z)$ to $M^2$.
Still, in the realistic case including RG effects beyond our toy model,
$C_2\ne 0$ and $C_T\ne 0$ will be reinstated at $\mu_R>M$, and 
$\Delta(\Lambda_h)$ with $\Lambda_h>M$ will sensibly capture the $\mu^2(m_Z)$ sensitivity to $M^2$.
A question remains as to whether $\Delta(M^+)$ acts sensibly in this scenario.
Is it possible to choose $\lambda_{H1,2}(M)$ such that $C_2(M)=0$ and $C_T(M)=0$
and the relevant terms $\propto M^2$ in \eq{EqDeltaToyModelF2} vanish?
Indeed, it is possible.
However, once RG effects are included, this is not sufficient to minimise $\Delta(M^+)$.
In particular, $d C_T/d \log\mu_R$ will generally be non-zero, leading to
an extra term in the $\partial/\partial\log M^2$ sensitivity measure:
\begin{align}
 \lim\limits_{\Lambda_h\to M^+} \left| \frac{\partial\log\mu^2(m_Z)}{\partial\log M^2} \right| = &\ 
  \frac{1}{2}\left| \frac{M^2}{\mu^2(m_Z)} \left( C_2(M) + 2C_T(M) + \left. \frac{d C_T}{d \log\mu_R}\right|_{\mu_R=M} \right) \right| .
  \label{EqDeltaToyModelF3}
\end{align}
The extra term can be thought of as arising from a shift in the matching scale.
If $\lambda_{H1}(M)=0$ is chosen so that $C_T(M)=0$ 
in order to minimise the $\partial/\partial\log \mu^2(\Lambda_h)$ sensitivity,
the full sensitivity measure is no longer minimised
for $C_2(M)=0$.
Instead, one would like to set $\left[ C_2(M) + d C_T/d \log\mu_R|_{\mu_R=M} \right] = 0$.
However, $d C_T/d \log\mu_R|_{\mu_R=M}$ is itself a function of $\lambda_{H1,2}$ (and gauge couplings),
thus it is not guaranteed that this is possible.
Indeed, in the cases we explore, it is not;
remarkably, $d C_T/d \log\mu_R|_{\mu_R=M} \supset +24 Q_3 \lambda_{H2}^2/(4\pi)^4$,
which cancels the negative contribution in \eq{EqC2SScalar2}
and leaves $\left[ C_2(M) + d C_T/d \log\mu_R|_{\mu_R=M}\right]$ positive
for any value of $\lambda_{H2}(M)$ when $\lambda_{H1}(M)=0$.\footnote{%
The reader might wonder if this is just a convenient happenstance.
It is possible.
However, we note that extending to three-loop RGEs with two-loop matching, 
this objection becomes moot.
At higher loop matching the threshold correction
will generally become a function of both $\lambda_{H1}$ and $\lambda_{H2}$.
The $\partial/\partial\log \mu^2(\Lambda_h)$ sensitivity is minimised for $C_T(M)=0$.
However, $\partial/\partial\log\lambda_{H1,2}(\Lambda_h)$ terms
also appear in the full sensitivity measure. The simultaneous vanishing of these terms is 
in general only guaranteed if $\lambda_{H1}(M)=\lambda_{H2}(M)=0$.
Plainly this restriction is too severe to absorb sensitivity arising elsewhere.}
Our numerical study captures this,
and we always recover a sensible value for $\Delta(M^+)$.

In any case, the possibility of a miraculous cancellation is clearly not the generic case,
and such cancellations are anyway quickly violated 
in the realistic scenario with RG effects and $\Lambda_h>M$.
Nevertheless we find it interesting that, 
in this framework (and even in the limit $\Lambda_h\to M^+$ limit),
there is a certain amount of $\mu^2(m_Z)$ sensitivity which cannot
be made to go away by a judicious choice of quartic couplings.

Much of the discussion here has only been of technical interest 
since we have chosen to project out our unknowns 
by minimising the sensitivity measure over them.
The reason for discussing 
possible regions of cancellation in such detail
is to highlight the fact that, with this sensitivity measure, 
there appears to be no place in parameter space to ``hide''. 
Where it appears that there might be
somewhere to hide, we have argued that 
apparent ``throats'' disappear at higher loop orders, 
or that sensitivity is always captured 
by some subcomponent of the sensitivity measure. 
It is important to establish this fact, 
since any violation would suggest that 
the sensitivity measure is incomplete,
and until one can establish that the sensitivity measure 
is always capturing the fine-tuning in these most simple of cases, 
then why would one trust it on more complicated models?

\subsection{Naturalness bounds}

\begin{figure}[t]
 \centering
 \includegraphics[width=0.6\columnwidth]{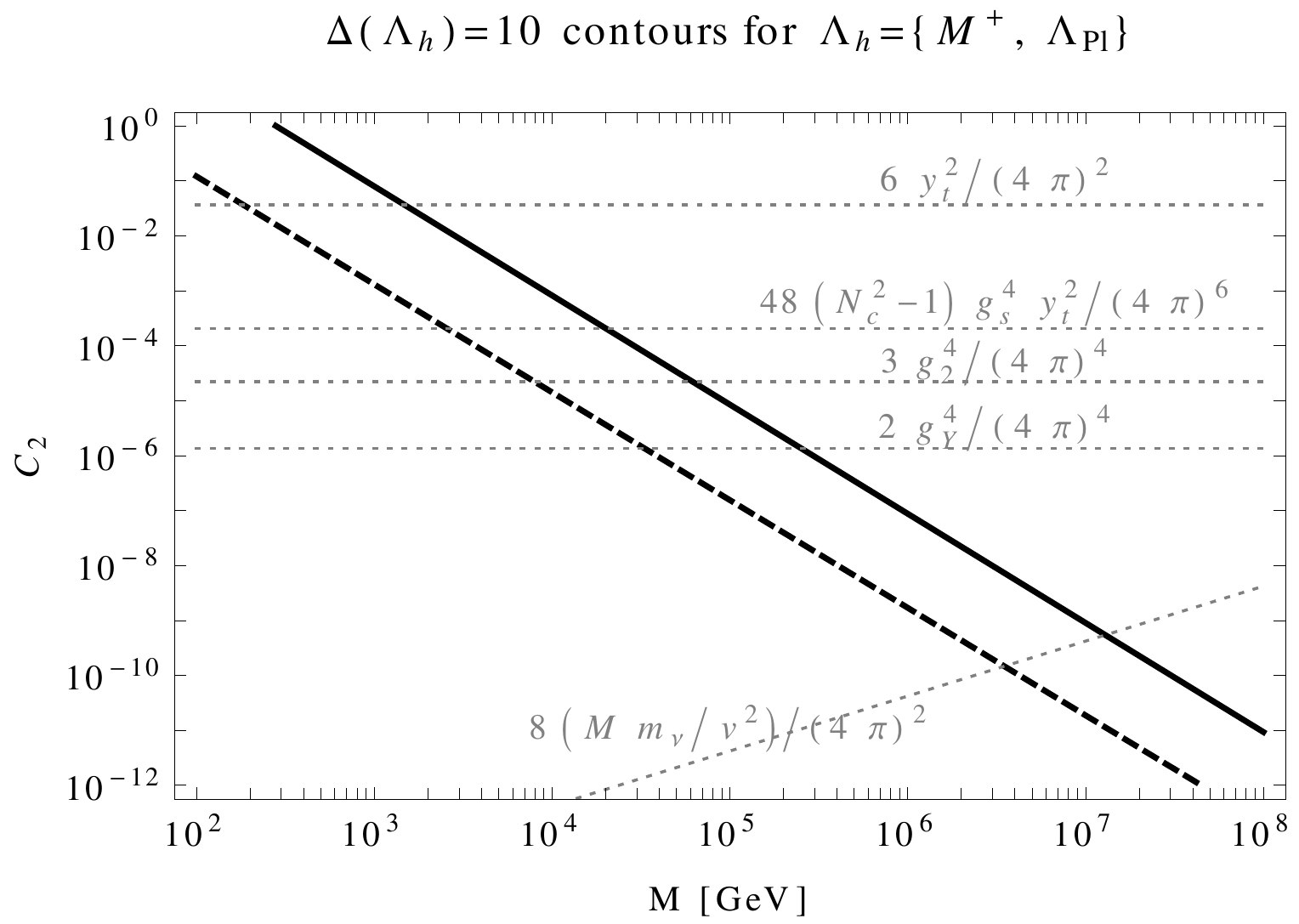}
 \caption{
 $\Delta(M,\Lambda_h)=10$ contours for $\Lambda_h=M^+$ (solid), or $\Lambda_{Pl}\sim 10^{19}$~GeV (dashed).
 Also shown as gray dotted lines are approximate $C_2$ contours for some benchmark heavy particles.
 The results for negative values of $C_2$ are very similar.}
 \label{figC2M}
\end{figure}

Naturalness bounds can be derived simply by bounding the sensitivity measure \eq{EqFTuning1}.
In \fig{figC2M} we show the $\Delta(\Lambda_h)=10$ contours
for $\Lambda_h=M^+$, or $\Lambda_{Pl}\sim 10^{19}$~GeV in our fermion-like toy model \eq{EqDeltaToyModelF}.
Points in parameter space below these lines can be considered natural,
and points above increasingly unnatural.

\fig{figC2M} can be used to estimate naturalness bounds 
on the masses of fermionic particles.
Consider for example a heavy fermion with a top-like coupling strength such that $C_2=6y_t^2/(4\pi)^2$;
taking $y_t^2=y_t^2(m_Z)\approx 0.96$ and reading across one finds a naturalness bound $M\lesssim$~TeV.
For a right-handed neutrino involved in a Type I see-saw, 
$C_2=4y_\nu^2/(4\pi)^2$ with $y_\nu^2 \simeq M m_\nu / (174\text{ GeV})^2$; 
taking $m_\nu=0.05$~eV results in a naturalness bound $M\lesssim 10^7$~GeV \cite{Vissani1997ys,Clarke2015gwa}.\footnote{%
This is not quite the correct thing to do, 
since the observable at low scale is $m_\nu$ 
and not $C_2$ [which was assumed to derive \eq{EqDeltaToyModelF2}]. 
Rest assured that using the appropriate sensitivity measure derived from \eq{EqFTuning0}
only marginally changes this picture.}
The reason that this naturalness bound is so large is simply because $C_2$ is so small.
Indeed, in the limit $C_2\to 0$ there is no naturalness bound on $M$.
In models with gauge singlets, 
$C_2\to 0$ can correspond to a technically natural limit \cite{Volkas1988cm,Foot2013hna} 
associated with decoupling of the particle from the SM fields.
It makes sense that there is no Higgs naturalness bound on the mass of such a particle,
given that in this limit the heavy particle can no longer ``talk'' to the Higgs at all.

The focus of this paper will be Higgs naturalness 
within the EFT of the SM plus a heavy GM. 
For fermionic GMs with $SU(3)$, $SU(2)$, or $U(1)_Y$ 
charge $Q_3$, $Q_2$, or $Q_1$,
the leading pure gauge contributions to $C_2$ are
$- 2 Q_1^2 g_1^4 / (4\pi)^4$,
$- \frac12 Q_2 (Q_2^2-1) g_2^4 / (4\pi)^4$,
and $+48 (N_C^2-1) g_3^4 y_t^2 / (4\pi)^6$, respectively.
Taking 
$g_1^2\approx 0.13$,
$g_2^2\approx 0.43$,
and $g_3^2\approx 1.48$,
these correspond to rough naturalness bounds of
(perhaps surprisingly to some)
tens to hundreds of TeV,
as sketched in \fig{figC2M}.
The size of the mass bounds is just a reflection of the smallness of $g^4/(4\pi)^4$.
The main purpose of this paper is to derive these bounds more rigorously;
we perform a full two-loop analysis
to examine the effects of adding various
(vector-like) fermionic and scalar GMs to the SM.
The above naturalness bound approximations turn out to be quite good 
for the fermionic GMs,
but they significantly deviate for scalar gauge multiplets,
since these always couple directly to the Higgs via a quartic term(s).
As already indicated, sensitivity to the RG evolved quartics must be properly taken into account.

\subsection{Comment on the Planck-weak hierarchy}

Before leaving this section, we want to comment on how 
the Planck-weak hierarchy fits into in this picture.
From \eqs{Eqmu2mZ} and (\ref{EqFTuning2}), one can see that $\Delta\simeq 1$ in the pure SM limit,
i.e. there is no enhanced sensitivity
when the Higgs mass is promoted to a high scale input parameter.
This should come as no surprise, 
since the only explicit scale in the SM is $\mu^2$ itself:
the value $\mu^2(\Lambda_h)$ is multiplicatively related to the value $\mu^2(m_Z)$
and remains electroweak scale up to high scales.\footnote{%
For the SM at two-loop we find $\mu^2(\Lambda_{Pl})\simeq -(94\text{ GeV})^2$ 
and $\Delta(\Lambda_{Pl})\simeq1$ to one part in $10^6$.}
Indeed, the effective Higgs potential remains consistent (albeit metastable)
even up to the very highest scale to which the 
SM can be valid: $\Lambda_{Pl}\sim 10^{19}$~GeV.
Now, it could be that gravity introduces 
large and physical corrections to $\mu^2(\mu_R)$ 
[or some related parameter(s)]
at or below this scale.
However, without a complete theory of quantum gravity,
we cannot calculate these corrections.\footnote{There exist extensions of Einstein gravity in which corrections to $\mu^2$ are both calculable and naturally small (see e.g.~\cite{Salvio:2014soa}), although such theories generally also have problems with unitarity.}
This picture therefore claims that
the SM with inputs at $\Lambda_{Pl}$ is natural,
in the sense that the low-energy observable $\mu^2(m_Z)$ 
is not extremely sensitive to the input $\mu^2(\Lambda_{Pl})$.
In such a case, one could sensibly ask:
why is $\mu^2(\Lambda_{Pl}) \ll \Lambda^2_{Pl}$?
We do not address this problem.
By construction our sensitivity measure remains agnostic to this input value by
assuming a flat prior in $\log\mu^2(\Lambda_{Pl})$.
Of course, as we have argued,
the presence of a heavy gauge multiplet
{\it can introduce a calculable and physical naturalness problem
irrespective of the situation with gravity},
and this is the problem that we study.
In such models a flat prior belief in $\log\mu^2(\Lambda_h)$
devolves to a low scale posterior belief which favours $\mu^2(m_Z)\gtrsim C_2 M^2$.
It could be that gravity behaves in a similar way,
but we cannot yet perform the calculation.

\section{Method \label{SecMethod}}

The main purpose of this paper is to derive and present naturalness bounds
on the masses of GMs within SM+GM EFTs valid up to scale $\Lambda_h$.
In \sec{SecNaturalness} we motivated a general procedure for determining these bounds:
take the low energy observables at $m_Z$,
evolve them under the RGEs to the scale $\Lambda_h$,
then evaluate and bound the sensitivity measure \eq{EqFTuning0}.
Presently we detail our method. 
We use sets of two-loop RGEs generated using a modified version of \textsc{PyR@TE} \cite{Lyonnet2013dna}.

The low scale observables are taken as the logarithms of 
SM $\overline{{\rm MS}}$ Lagrangian parameters at scale $m_Z$:
$\exp(\mathcal{O}_i)=\{ \mu^2(m_Z),$ $\lambda(m_Z),$ $g_1(m_Z),$ $g_2(m_Z),$ $g_3(m_Z),$ $y_t(m_Z),$ $y_b(m_Z),$ 
$y_\tau(m_Z) \}=\{ -(88\text{ GeV})^2,$ $0.13,$ $0.36,$ $0.66,$ $1.22,$ $0.96,$ $0.017,$ $0.010 \}$.
For simplicity we ignore the Higgs and the top quark thresholds.
The high scale input parameters are taken as the
logarithms of the minimal set of SM+GM $\overline{{\rm MS}}$ Lagrangian parameters at scale $\Lambda_h$
(to be explicitly listed in the following subsections);
by minimal we mean that terms in the SM+GM Lagrangian 
which can be set to zero in a technically natural way are not included.
The observables are numerically evolved under the two-loop SM RGEs up to 
the threshold of the GM, $\mu_R=M$,
where we perform one-loop matching onto the parameters of the SM+GM EFT.
The mass parameter for the GM is also a renormalised $\overline{\rm MS}$ parameter, 
which we set equal to $M$ at the scale $\mu_R=M$.\footnote{The 
bounds we present are therefore bounds on the parameter $M$
(i.e. the $\overline{\rm MS}$ mass of the GM 
at the scale $M$),
not the pole mass.} 
New parameters are introduced in the case of a scalar GM; 
these are left as free parameters which are
numerically minimised over when evaluating the sensitivity measure.
The two-loop SM+GM RGEs are used
to evolve all parameters up to the high scale $\Lambda_h$.
The approximation to the full sensitivity measure, \eq{EqFTuning2}, 
is then evaluated numerically 
by varying the appropriate input parameters around their values at $\Lambda_h$,
evolving all parameters back down to the scale $m_Z$,
matching the SM+GM EFT onto the SM EFT at the matching scale $\mu_R$ given by $M(\mu_R)=\mu_R$,
and measuring the change in the Higgs mass parameter.

\subsection{Vector-like fermion}

The minimal SM+GM Lagrangian for a vector-like fermion is that of the SM plus
\begin{align}
 \Delta\mathcal{L} = \bar{\psi}D^\nu\gamma_\nu\psi - M\bar{\psi}\psi .
\end{align}
The high scale input parameters of this model
are those of the SM plus the renormalised parameter $M(\Lambda_h)$,
i.e. $\mathcal{I}_j=\{ \mu^2(\Lambda_h),$ $\lambda(\Lambda_h),$ 
$g_1(\Lambda_h),$ $g_2(\Lambda_h),$ $g_3(\Lambda_h),$ $y_t(\Lambda_h),$ $y_b(\Lambda_h),$ $y_\tau(\Lambda_h)$, $M(\Lambda_h) \}$.
The one-loop matching conditions are trivial,
\begin{align}
 \mu^2_+(\mu_R) = \mu^2_-(\mu_R) \ , &&
 \lambda_+(\mu_R) = \lambda_-(\mu_R) \ ,
\end{align}
where the $+$ ($-$) subscript denotes the SM+GM (SM) EFT parameter.

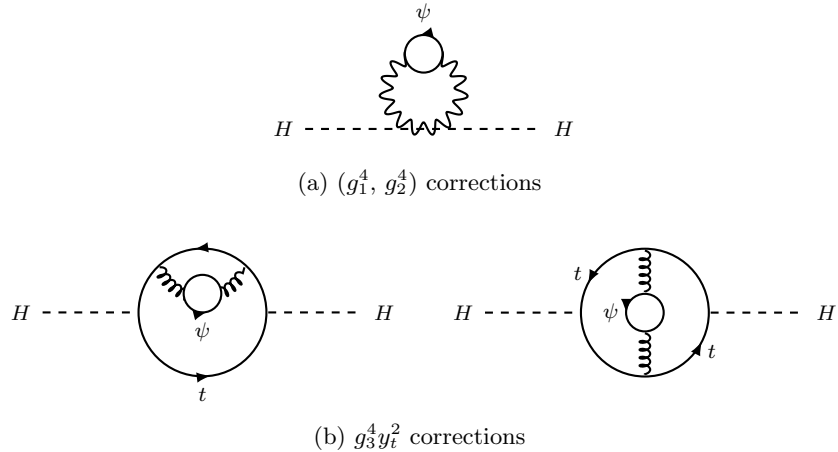
\begin{figure}[t]
  \centering
  \captionsetup[subfigure]{}
  \subfloat[($g_1^4$, $g_2^4$) corrections \label{FigFermionCorrectionsA}]{
    \centering
    \iftikz    
    \begin{tikzpicture}
      \scriptsize
      \node[vtx, label=180:$H$] (i1) at (0.6,0) {};
      \node[vtx, label=0:$H$] (o1) at (3.8,0) {};
      \node[loop] (l1) at (2.2,0.5) {};    
      \node[loop] (l2) at (2.2,1.0) {};    
      \graph[use existing nodes]{
      i1 --[scalar] o1;
      };  
      \draw[photonloop] (l1) circle(0.5);      
      \draw[fermion, preaction={fill,white}, decoration={reset marks, markings, mark=at position 1/4+0.04 with {\arrow[]{latex}}}] (l2) circle(0.25);      
      \node at (2.2,1.55) {$\psi$};
    \end{tikzpicture} 
    \fi
  }\\
  \subfloat[$g_3^4y_t^2$ corrections \label{FigFermionCorrectionsB}]{
    \centering
    \iftikz    
    \begin{tikzpicture}
      \scriptsize
      \node[vtx, label=180:$H$] (i1) at (0.4,0) {};
      \node[vtx] (v1) at (1.7,0) {};
      \node[vtx] (v2) at (3.4,0) {};
      \node[vtx] (v3) at (1.98,0.6) {};     
      \node[vtx, label=0:$H$] (o1) at (4.7,0) {};
      \node[loop, outer sep=7pt] (l1) at (2.55,0) {};    
      \node[loop] (l3) at (2.55,0.25) {};      
      \graph[use existing nodes]{
      i1 --[scalar] v1;
      v2 --[scalar] o1;     
      };  
      \draw[fermion, decoration={reset marks, markings, mark=between positions 1/4.+0.02 and 1 step 1/2. with {\arrow[]{latex}}}] (l1) circle(0.85);       
      \draw[gluon, decoration={coil, amplitude=2pt, segment length=3pt}] (v3) arc(200:340:0.6);      
      \draw[fermion, preaction={fill,white}, decoration={reset marks, markings, mark=at position 3/4+0.04 with {\arrow[]{latex}}}] (l3) circle(0.25);         
      \node at (node cs:name=l1,angle=270) {$t$};      
      \node at (2.55,-0.25) {$\psi$};        
    \end{tikzpicture}
    \hspace{1em}
    \begin{tikzpicture}
      \scriptsize
      \node[vtx, label=180:$H$] (i1) at (0.4,0) {};
      \node[vtx] (v1) at (1.7,0) {};
      \node[vtx] (v2) at (3.4,0) {};
      \node[vtx] (v3) at (2.55,0.85) {};
      \node[vtx] (v4) at (2.55,0.25) {};
      \node[vtx] (v5) at (2.55,-0.25) {};
      \node[vtx] (v6) at (2.55,-0.85) {};
      \node[vtx, label=0:$H$] (o1) at (4.7,0) {};
      \node[loop, outer sep=5pt] (l1) at (2.55,0) {};
      \node at (node cs:name=l1,angle=150) {$t$};
      \node at (node cs:name=l1,angle=330) {$t$};
      \node at (2.1,0) {$\psi$};         
      \graph[use existing nodes]{
      i1 --[scalar] v1;
      v2 --[scalar] o1;
      v3 --[gluon, decoration={coil, amplitude=2pt, segment length=3pt}] v4;
      v5 --[gluon, decoration={coil, amplitude=2pt, segment length=3pt}] v6;
      };  
      \draw[fermion, decoration={reset marks, markings, mark=between positions 3/8.+0.05 and 1 step 1/2. with {\arrow[]{latex}}}] (l1) circle(0.85);
      \draw[fermion, decoration={reset marks, markings, mark=at position 1/2. with {\arrow[]{latex}}}] (l1) circle(0.25);            
      \node at (2.9,-1.18) {};      
    \end{tikzpicture}
    \fi
  }
  \caption{Corrections to $\mu^2$ from a heavy vector-like fermion.}
  \label{FigFermionCorrections}
\end{figure}

The $\mu^2(\mu_R)$ RGE takes the form of \eq{EqRGE} 
with $C_2(\mu_R)$ a function of SM parameters.
Recall that it is primarily the $C_2(\mu_R)$ term which leads to a potential naturalness problem,
as argued in \sec{SecExample} for constant $C_2$.
In the vector-like fermionic SM+GM EFT it takes the form
\begin{align}
 C_2 = 
  - 2 Q_3 Q_2 Q_1^2 \frac{g_1^4}{(4\pi)^4} 
  - \frac12 Q_3 Q_2 (Q_2^2-1) \frac{g_2^4}{(4\pi)^4} 
  + 96 Q_2 (N_c^2-1) c(r_\psi) \frac{g_3^4 y_t^2}{(4\pi)^6},
\end{align}
where $c(r_\psi)=\frac12$ $(3)$ for $Q_3=3\,(8)$,
and dependence on scale $\mu_R$ is implied.
Representative diagrams leading to these terms are shown in \fig{FigFermionCorrections}.
Note that we have added by hand the leading three-loop $SU(3)$ correction,
arising from three diagrams [see \fig{FigFermionCorrectionsB}],
since otherwise the two-loop RGEs do not capture any $SU(3)$ correction beyond multiplicity factors.\footnote{This
three-loop correction was calculated with the aid of \textsc{Matad} \cite{Steinhauser2000ry}.}
This correction turns out to be competitive with the two-loop pure gauge corrections
at scales $\mu_R \lesssim10^5$~GeV due to the relatively large couplings $g_3$ and $y_t$ below this scale.
It is also opposite in sign, 
thus potentially delaying the growth of $\mu^2(\mu_R)$ (and the corresponding naturalness problem)
if it happens to approximately cancel with the other gauge contributions.

There are no unconstrained high scale dimensionless inputs to minimise over,
so the sensitivity measure \eq{EqFTuning2} is just
\begin{align}
  \Delta(M,\Lambda_h)
= \sqrt{ \left(\frac{\partial\log\mu^2(m_Z)}{\partial\log\mu^2(\Lambda_h)}\right)^2 +
         \left(\frac{\partial\log\mu^2(m_Z)}{\partial\log M(\Lambda_h)}\right)^2 }.
  \label{EqFTuningFermion}
\end{align}
Results are obtained by numerically evaluating $\Delta(M,\Lambda_h)$ at points of interest in $(M,\Lambda_h)$ space.

\subsection{Complex scalar \label{SecMethodScalar}}

The minimal SM+GM Lagrangian for a complex scalar is that of the SM plus
\begin{align}
 \Delta\mathcal{L} = D^\nu \Phi^\dagger D_\nu \Phi - \left(
  M^2 \Phi^\dagger \Phi + 
  \sum \lambda_{\Phi i} \Phi^\dagger \Phi \Phi^\dagger \Phi+ 
  \sum \lambda_{H j} H^\dagger H \Phi^\dagger \Phi \right),
\end{align}
where $H$ is the SM Higgs field,
and the sums are over all possible contractions.
Explicitly, we take the following convenient contractions for the portal quartics 
\begin{equation}
 \Delta\mathcal{L}\supset\lambda_{H1}H^\dagger H \,\mathrm{Tr}(\Phi^\dagger\Phi)
 + \lambda_{H2}\left(2\,\mathrm{Tr}(H^\dagger\Phi\Phi^\dagger H)-H^\dagger H \,\mathrm{Tr}(\Phi^\dagger\Phi)\right),
\end{equation}
where the second term is relevant only for $Q_2\geq2$.
The high scale input parameters of the model 
are those of the SM plus the extra self and portal quartics and renormalised mass parameter,
i.e. $\mathcal{I}_j=\{ \mu^2(\Lambda_h),$ $\lambda(\Lambda_h),$ 
$g_1(\Lambda_h),$ $g_2(\Lambda_h),$ $g_3(\Lambda_h),$ $y_t(\Lambda_h),$ $y_b(\Lambda_h),$ $y_\tau(\Lambda_h),$
$\lambda_{\Phi 1}(\Lambda_h),\dots,\lambda_{H 1}(\Lambda_h),\dots , M^2(\Lambda_h)\}$.
The one-loop matching conditions are 
\begin{align}
 \mu^2_+(\mu_R) &= \mu^2_-(\mu_R) - Q_3 Q_2 \frac{\lambda_{H1}(\mu_R)}{(4\pi)^2} M^2(\mu_R) \left[1-\log\left(\frac{M^2(\mu_R)}{\mu_R^2}\right)\right] \ ,\\
 \lambda_+(\mu_R) &= \lambda_-(\mu_R) - \frac{Q_3 Q_2}{2} \frac{\lambda_{H1}^2(\mu_R)}{(4\pi)^2}  \log\left(\frac{M^2(\mu_R)}{\mu_R^2}\right)  \ ,
\end{align}
where the $+$ ($-$) subscript denotes the SM+GM (SM) EFT parameter and we have neglected terms suppressed by powers of $v^2/M^2$.
We will always work in the limit where $\Phi$ does not obtain a vacuum expectation value.
This is a well-motivated simplification, since for masses at the naturalness bounds we will obtain (typically $M>$~TeV),
experimental agreement with the 
canonical Higgs mechanism for electroweak symmetry breaking
generically constrains any scalar GM to observe this limit,
and of course a coloured scalar multiplet must exactly satisfy it.
Evidently the $\mu^2$ term receives a threshold correction 
when matching is performed at the scale $\mu_R=M(\mu_R)$.

\begin{figure}[t]
  \centering
  \captionsetup[subfigure]{}
  \subfloat[($g_1^4$, $g_2^4$) and $\lambda_{Hi}$ corrections to $\mu^2$ \label{FigScalarCorrectionsA}]{
    \centering
    \iftikz    
    \begin{tikzpicture}
      \scriptsize
      \node[vtx, label=180:$H$] (i1) at (0.6,0) {};
      \node[vtx, label=0:$H$] (o1) at (3.8,0) {};
      \node[loop] (l1) at (2.2,0.5) {};
      \node[loop] (l2) at (2.2,1.5) {};      
      \graph[use existing nodes]{
      i1 --[scalar] o1;
      };  
      \draw[photonloop] (l1) circle(0.5);
      \draw[scalar] (l2) circle(0.5);    
      \node at (2.2,1.75) {$\Phi$};         
    \end{tikzpicture} 
    \hspace{1em}    
    \begin{tikzpicture}
      \scriptsize
      \node[vtx, label=180:$H$] (i1) at (0.6,0) {};
      \node[vtx, label=0:$H$] (o1) at (3.8,0) {};
      \node[loop] (l1) at (2.2,0.5) {};    
      \graph[use existing nodes]{
      i1 --[scalar] o1;
      };  
      \draw[scalar] (l1) circle(0.5);  
      \node at (2.2,0.75) {$\Phi$};       
    \end{tikzpicture}   
    \fi
  }\\
  \subfloat[($g_1^4$, $g_2^4$, $g_1^2g_2^2$) and $g_3^4y_t^2$ corrections to $\lambda_{Hi}$ \label{FigScalarCorrectionsB}]{
    \centering
    \iftikz    
    \begin{tikzpicture}
      \scriptsize
      \node[vtx, label=180:$\Phi$] (i1) at (0.4,0.5) {};
      \node[vtx, label=180:$\Phi$] (i2) at (0.4,-0.5) {};      
      \node[vtx] (v1) at (1.7,0) {};
      \node[vtx] (v2) at (2.7,0) {};
      \node[vtx, label=0:$H$] (o1) at (3.8,0.5) {};
      \node[vtx, label=0:$H$] (o2) at (3.8,-0.5) {};      
      \node[loop] (l1) at (2.2,0) {};      
      \graph[use existing nodes]{
      i1 --[scalar] v1 --[scalar] i2;
      o1 --[scalar] v2 --[scalar] o2;
      };  
      \draw[photonloop] (l1) circle(0.5);      
    \end{tikzpicture} 
    \hspace{1em}    
    \begin{tikzpicture}
      \scriptsize
      \node[vtx, label=180:$\Phi$] (i1) at (0.6,0.5) {};
      \node[vtx, label=180:$\Phi$] (i2) at (0.6,-0.5) {};      
      \node[vtx] (v1) at (1.7,0) {};
      \node[vtx] (v2) at (2.7,0.5) {};
      \node[vtx] (v3) at (2.7,-0.5) {};      
      \node[vtx] (v4) at (3.7,-0.5) {};
      \node[vtx] (v5) at (3.7,0.5) {};      
      \node[vtx, label=0:$H$] (o1) at (4.8,0.5) {};
      \node[vtx, label=0:$H$] (o2) at (4.8,-0.5) {};                
      \graph[use existing nodes]{
      i1 --[scalar] v1 --[scalar] i2;
      v2 --[fermion] v3 --[fermion] v4 --[fermion] v5 --[fermion] v2;
      v2 --[gluon] v1;
      v3 --[gluon] v1;
      o1 --[scalar] v5;
      o2 --[scalar] v4;
      };  
      \node at (2.9,0) {$t$}; 
      \node at (2.9,-0.75) {};
    \end{tikzpicture}  
    \fi
  }
  \caption{Corrections to $\mu^2$ from a heavy scalar, and related corrections to $\lambda_{Hi}$.}
  \label{FigScalarCorrections}
\end{figure}
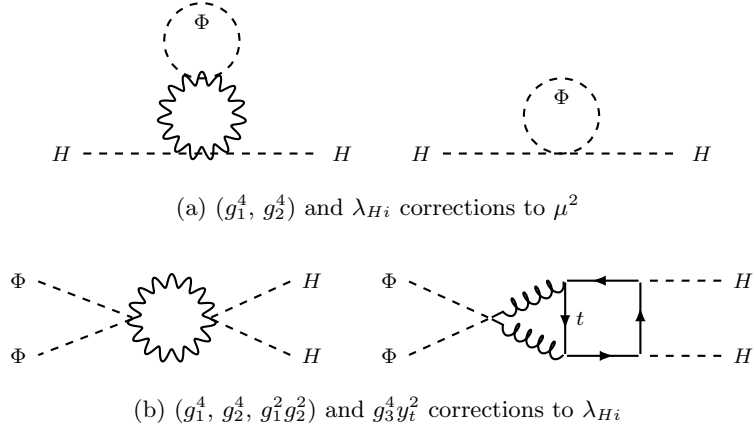

The $\mu^2(\mu_R)$ RGE in the SM+GM EFT with a complex scalar takes the form of \eq{EqRGE} 
with $C_2(\mu_R)$ a function of both the SM parameters and the extra quartics,
\begin{align}
 C_2 = & \ 
  + 2 Q_3 Q_2 \frac{\lambda_{H1}}{(4\pi)^2}
  - 4 Q_3 Q_2 \frac{\lambda_{H1}^2}{(4\pi)^4} \nonumber \\
 &+ 5 Q_3 Q_2 Q_1^2 \frac{g_1^4}{(4\pi)^4} 
  + \frac54 Q_3 Q_2 (Q_2^2-1) \frac{g_2^4}{(4\pi)^4} 
 \nonumber \\
 &+ 16 Q_3 Q_2 Q_1^2 \frac{g_1^2 \lambda_{H1}}{(4\pi)^4}
  + 4 Q_3 Q_2 (Q_2^2-1) \frac{g_2^2\lambda_{H1}}{(4\pi)^4}
 \nonumber \\
 & - 24 Q_3 \frac{\lambda_{H2}^2}{(4\pi)^4}
+ 64 Q_2 \frac{g_3^2\lambda_{H1}}{(4\pi)^4} \ ,
 \label{EqC2Scalar}
\end{align}
where the final line
$\lambda_{H2}^2$ ($g_3^2\lambda_{H1}$) term appears only if $Q_2=2,3$ ($Q_3=3$),
and dependence on scale $\mu_R$ is implied.\footnote{%
We encountered difficulties with \textsc{PyR@TE} when generating the two-loop scalar octet RGEs.
Thus, regrettably, they are left out of this study.}
Representative diagrams which lead to these terms are shown in \fig{FigScalarCorrectionsA}.
Recall that the naturalness problem can be ameliorated in the limit $C_2\to 0$.
Indeed, it could be the case that at some scale the $\lambda_{Hi}$ conspire to give $C_2(\mu_R)=0$.
However, this will not be stable under RG evolution;
the portal quartics receive gauge corrections via the diagrams shown in \fig{FigScalarCorrectionsB}:
\begin{align}
  \frac{d\lambda_{H1}}{d\log\mu_R} &\supset 3Q_1^2\frac{g_1^4}{(4\pi)^2} + 3c_2(r_\Phi)\frac{g_2^4}{(4\pi)^2} - 32\frac{y_t^2g_{3}^{4}}{(4\pi)^4}\ , \\
  \frac{d\lambda_{H2}}{d\log\mu_R} &\supset 3n_2Q_1\frac{g_1^2g_2^2}{(4\pi)^2} \ .   
\end{align}
Here, $n_2=1\,(2)$ and the quadratic Casimir $c_2(r_\Phi)=3/4\,(2)$ for $Q_2=2\,(3)$; and the term
proportional to $g_3$ applies to the case $Q_3=3$.
Note that there exist corrections with odd power in $Q_1$;
this means that (unlike the fermion case)
RG evolution will depend on the sign of the hypercharge,
however we do not observe any noticeable consequences from this effect.
Also, in this case, we see that the two-loop scalar RGEs 
{\it do} capture an $SU(3)$ correction $\sim g_3^4 y_t^2$, 
through a two-loop correction to the portal quartics.
Thus we do not add by hand the three-loop $g_3^4 y_t^2$ term 
to the $\mu^2$ RGE in our scalar SM+GM analysis.

The portal and self quartics are unknown and unconstrained 
high scale input parameters which must be projected out
to obtain a sensitivity measure which is a function of $(M,\Lambda_h)$.
Our measure \eq{EqFTuning1} requires them 
to take on such values which minimise the Bayes factor,
i.e. such values which give a conservative ``best case scenario'' 
for Higgs mass sensitivity in the given model.
In \eq{EqFTuning2}, 
we wrote down an approximation to the full sensitivity measure \eq{EqFTuning0},
which is valid when the low scale dimensionless SM observables
are approximately insensitive to
the unconstrained inputs in the vicinity of the minimum.
In that case we can evaluate the sensitivity measure \eq{EqFTuning2},
\begin{align}
 \Delta(M,\Lambda_h)
  = 
  \left.
  \min\limits_{\lambda_{Hi}} 
  \left\{  
  \sqrt{ 
  \begin{align}
  &\left( \frac{\partial\log\mu^2(m_Z)}{\partial\log\mu^2(\Lambda_h)} \right)^2 +
  \left( \frac{\partial\log\mu^2(m_Z)}{\partial\log M^2(\Lambda_h)} \right)^2 \\
  &+ \sum \left( \frac{\partial\log\mu^2(m_Z)}{\partial\log\lambda_{Hi}(\Lambda_h)} \right)^2 +
  \sum \left( \frac{\partial\log\mu^2(m_Z)}{\partial\log\lambda_{\Phi j}(\Lambda_h)} \right)^2
  \end{align}
  }
  \right|_{^{\lambda_{\Phi j}(M)=0}_{\lambda_{Hi}}}
  \right\} \ ,
  \label{EqFTuningScalar}
\end{align}
where $\lambda_{\Phi j}(M) = 0$ is minimally consistent with our assumption 
that $\Phi$ attains no vacuum expectation value.\footnote{%
The full sensitivity measure \eq{EqFTuningScalar} 
should also involve a minimisation over the $\lambda_{\Phi j}$.
However, we found that, after demanding $\lambda_{\Phi j}(M) \ge 0$ to ensure no non-trivial vacuum expectation value, 
the minimum always occurred for $\lambda_{\Phi j}(M) \simeq 0$.
Thus in practice, to improve speed and numerical stability,
we set $\lambda_{\Phi j}(M)=0$ when evaluating the sensitivity measure
and note that even varying this up to $\lesssim 0.5$ made little difference to our results.
We also note that the $\lambda_{\Phi j}$ always evolve to positive values due to 
pure gauge contributions to their RGEs at one-loop,
and therefore the potential does not become trivially unstable.}
We will now make an argument for why we indeed expect this approximation to be valid in the scalar SM+GM.

Intuitively, since $C_2 M^2$ is the primary quantity which leads to a naturalness problem,
to zeroth approximation we expect the minimum to occur where the
average value of $C_2(\mu_R)$ over the RG evolution is zero.
Consider the case with only one portal coupling $\lambda_{H1}$.
Then from \eq{EqC2Scalar} our expectation requires $\lambda_{H1}(\mu_R)$ 
to take on values $\mathcal{O}(g^4/(4\pi)^2)$
(in order to cancel the pure gauge contribution),
and for $C_2(\mu_R)$ to swap sign along its RG evolution.
Indeed, we observe this to be the case in our numerical study, as we will demonstrate in \sec{SecDiscussion}.
Now, $\lambda_{H1}$ enters the one-loop dimensionless SM parameter RGEs only for the Higgs self-quartic $\lambda$,
as $\sim \lambda_{H1}^2/(4\pi)^2$.
Therefore its contribution to the evolution of $\lambda$ (and all dimensionless SM parameters) is very small at the minimum,
and we can say that $\partial\lambda(m_Z)/\partial\lambda_{H1}(\Lambda_h)\simeq 0$.
As for the new quartic couplings $\lambda_{\Phi j}$, 
they do not directly enter any of the dimensionless SM parameter two-loop RGEs,
therefore their effect is also very small.
Extending to the case with two portal quartics,
\eq{EqC2Scalar} clearly implies that a contour in $(\lambda_{H1},\lambda_{H2})$ space
will satisfy $C_2(\mu_R)=0$, so our argument for $\mathcal{O}(g^4/(4\pi)^2)$ quartics no longer holds.
However the $\partial/\partial\log\lambda_{Hi}(\Lambda_h)$ terms in the sensitivity measure \eq{EqFTuningScalar}
are proportional to $\lambda_{Hi}(\Lambda_h)$,
so that the minimum will always prefer smaller values for these quartics.
The dimensionless SM observables will then be insensitive
to variations around $\lambda_{Hi}(\Lambda_h)$ for the reasons already argued.

\section{Results \label{SecResults}}

\begin{figure}[p]
 \begin{minipage}[c][0.95\textheight]{\textwidth}
  \centering
  \includegraphics[width=0.99\textwidth]{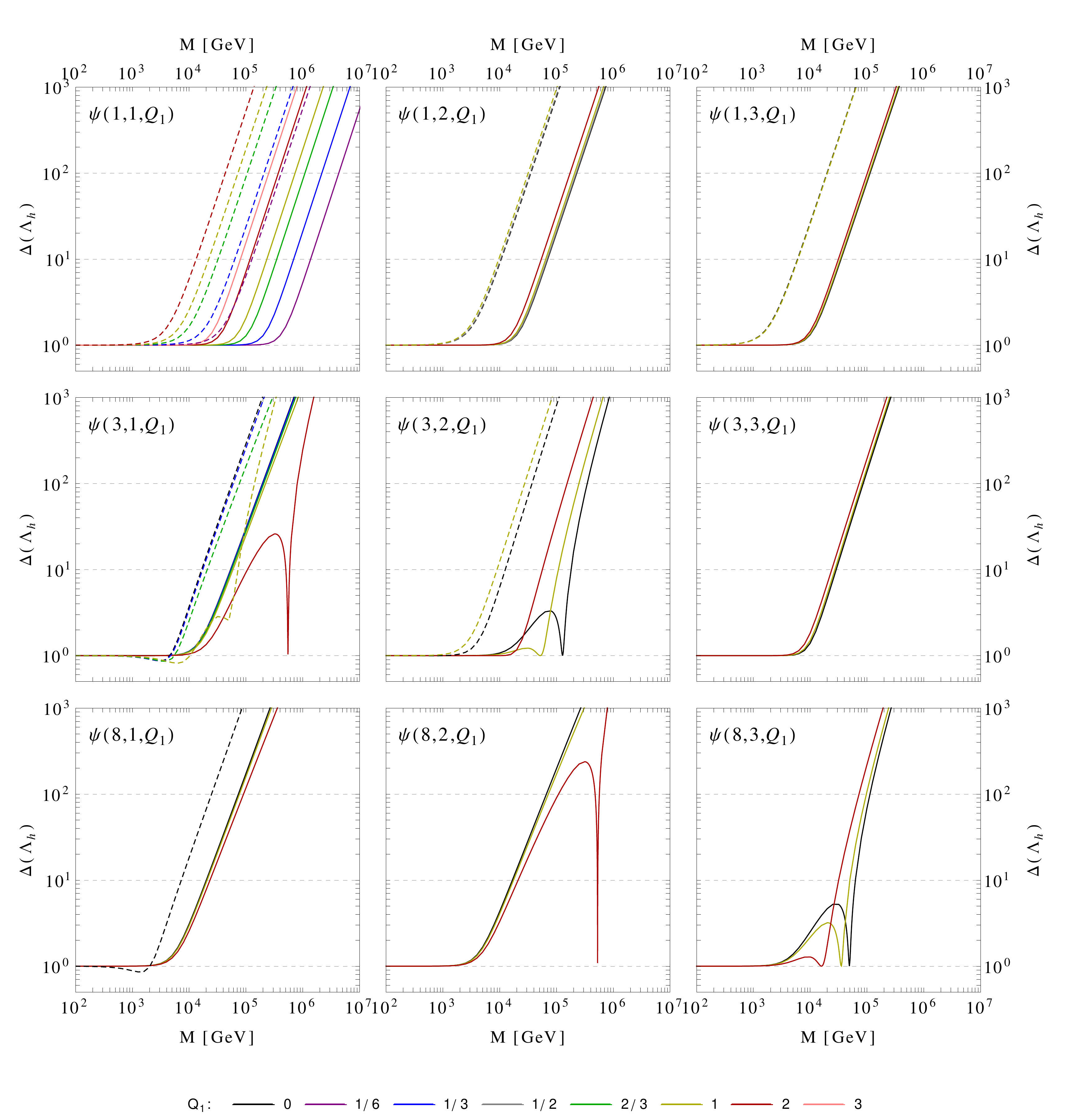}
  \caption{$\Lambda_h=\{ M^+, \Lambda_{Pl} \}$ contours \{solid, dashed\} 
  in the vector-like fermionic SM+GM EFT.
  The ``throat'' features are an artifact of the loop level to which we are working [see \sec{Sec2Fermion}].}
  \label{FigFermionDelta}
 \end{minipage}
\end{figure}

\begin{figure}[p]
 \begin{minipage}[c][0.95\textheight]{\textwidth}
  \centering
  \includegraphics[width=0.99\textwidth]{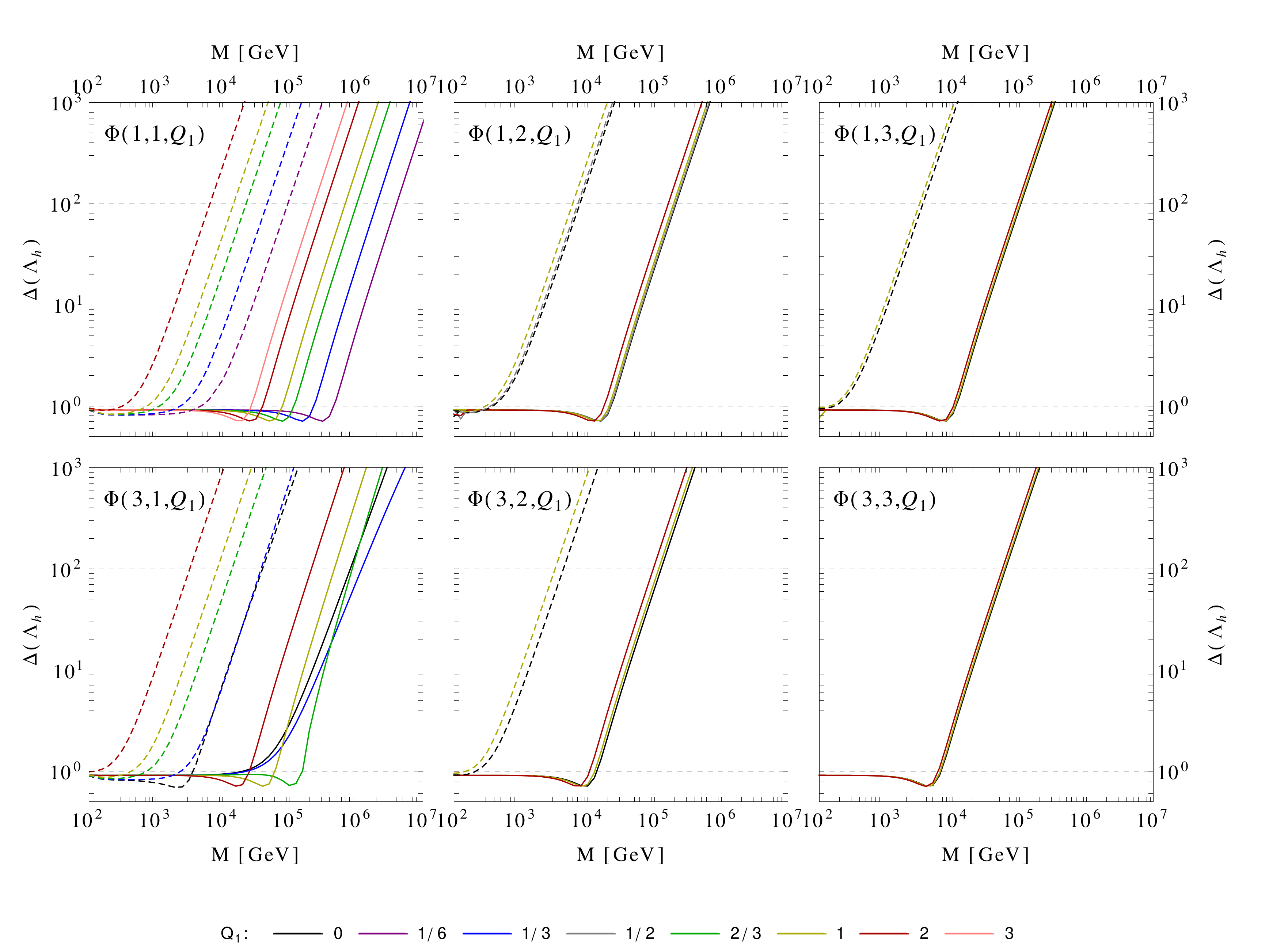}
  \caption{$\Lambda_h=\{ M^+, \Lambda_{Pl} \}$ contours \{solid, dashed\} 
  in the scalar SM+GM EFT.}
  \label{FigScalarDelta}
 \end{minipage}
\end{figure}

\begin{figure}[p]
 \begin{minipage}[c][0.95\textheight]{\textwidth}
  \centering
  \includegraphics[width=0.99\textwidth]{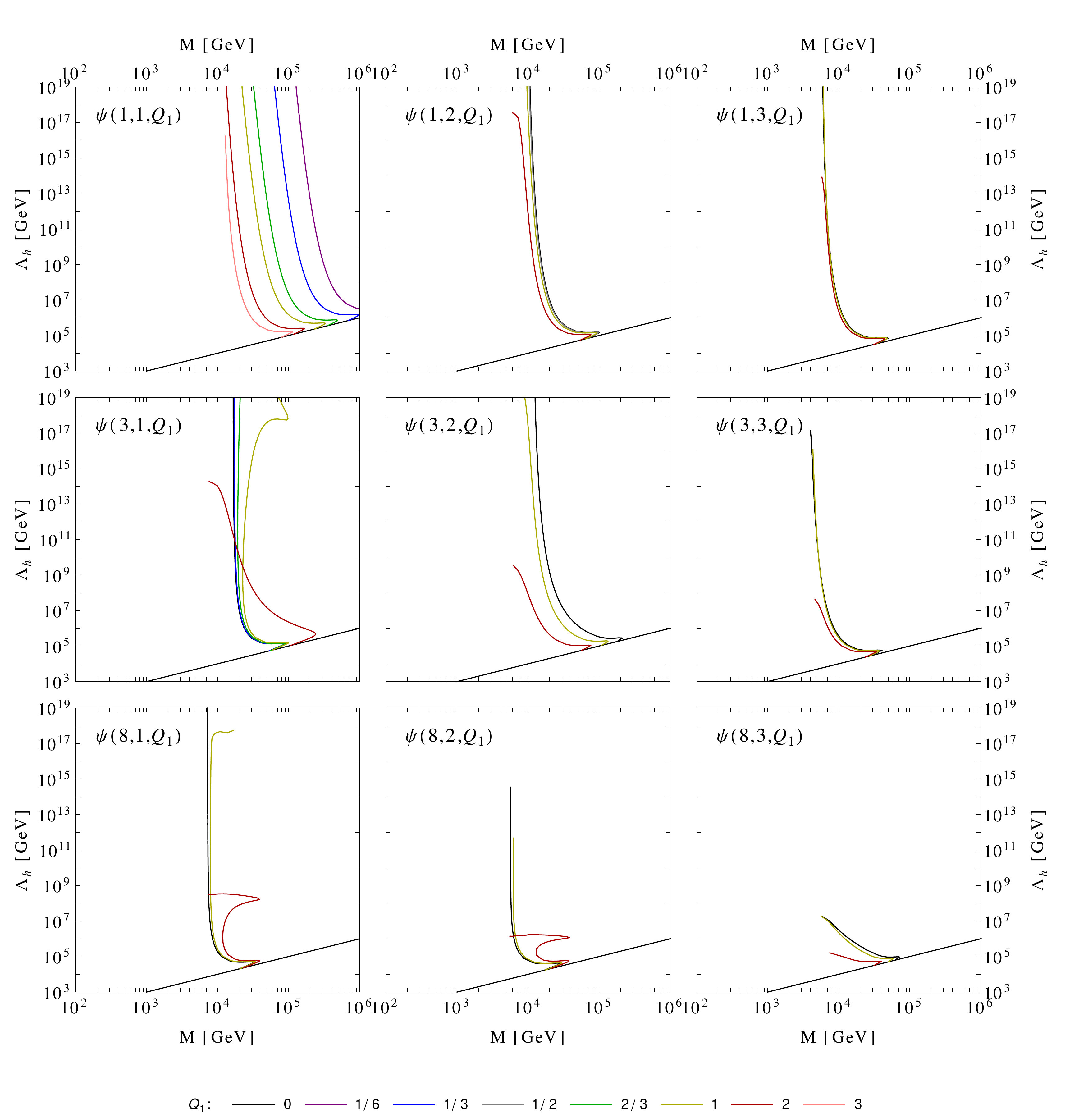}
  \caption{$\Delta(\Lambda_h)=10$ contours in the vector-like fermionic SM+GM EFT.
  If a line ends it is because the system hits a Landau pole.}
  \label{FigFermionLambda}
 \end{minipage}
\end{figure}

\begin{figure}[p]
 \begin{minipage}[c][0.95\textheight]{\textwidth}
  \centering
  \includegraphics[width=0.99\textwidth]{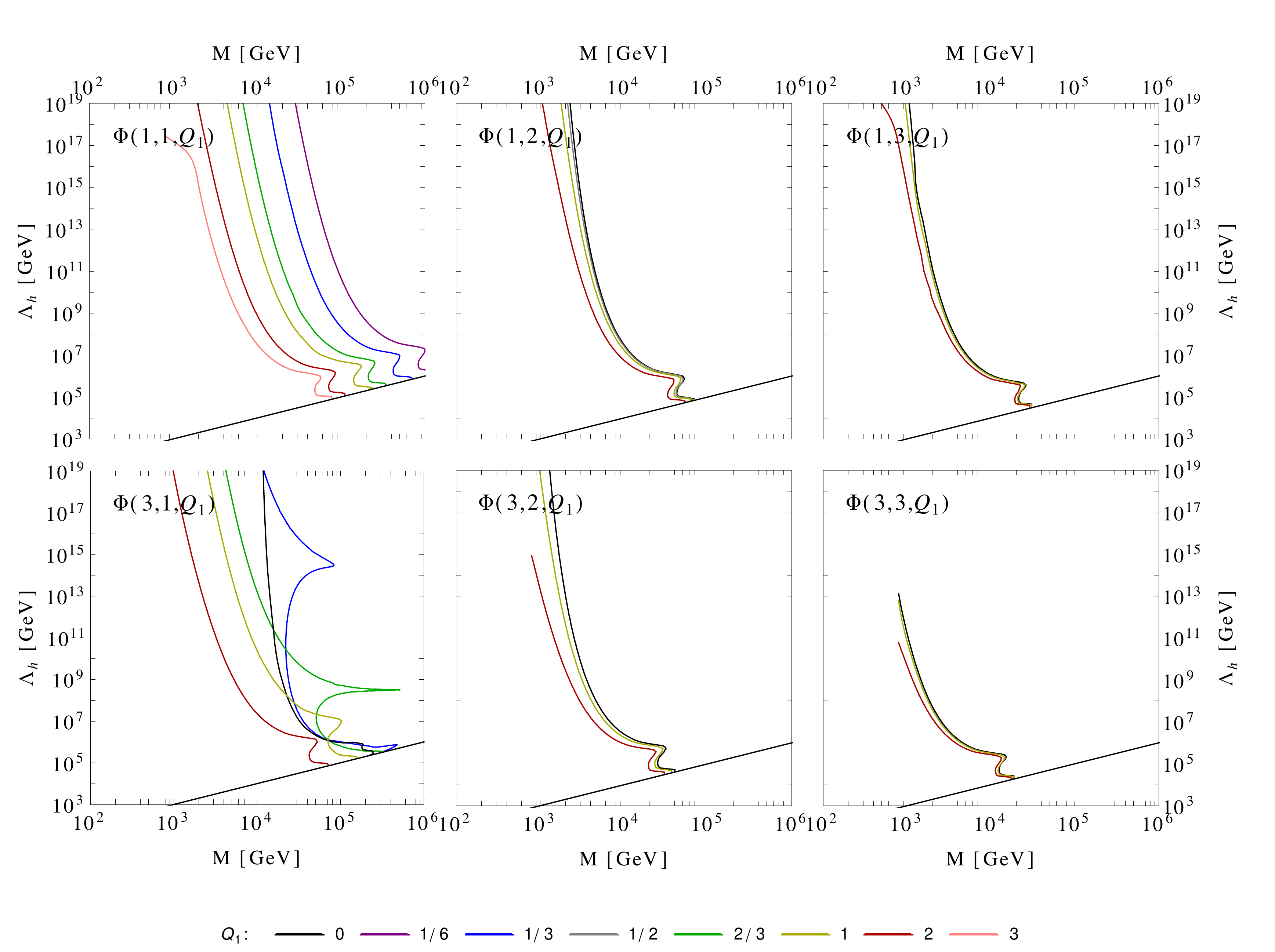}
  \caption{$\Delta(\Lambda_h)=10$ contours in the scalar SM+GM EFT.
  If a line ends it is because the system hits a Landau pole.
  }
  \label{FigScalarLambda}
 \end{minipage}
\end{figure}

The naturalness bounds for various vector-like fermionic and scalar GMs,
derived according to the method detailed in \sec{SecMethod},
are presented in \tabs{TabFermions} and \ref{TabScalars} for
$\Lambda_h=\{M^+,\Lambda_{Pl}\}$ and
$\Delta(\Lambda_h)=\{10,100,1000\}$.
Contour plots in $(M,\Delta)$ and $(M,\Lambda_h)$
parameter space are also provided in 
\figs{FigFermionDelta}--\ref{FigScalarLambda}.
These constitute the main result of this paper,
and we hope that they will serve as a useful reference of 
naturalness benchmarks for phenomenological model builders.

\begin{table}[p]
 \begin{minipage}[c][0.95\textheight]{\textwidth}
 \scriptsize
 \begin{tabular}{||ccc||c|c|c||c|c|c||}
 \hline
         &           &          & \multicolumn{3}{c||}{$\Lambda_h=M^+$}        & \multicolumn{3}{c||}{$\Lambda_h=10^{19}$~GeV} \\
 \hline
 $SU(3)$ & $SU(2)_L$ & $U(1)_Y$ & $\Delta=10$ 	& $\Delta=100$ 	& $\Delta=1000$ & $\Delta=10$ 	& $\Delta=100$ 	& $\Delta=1000$ \\
 \hline\hline
 1&1&$\pm\frac16$		& 1400 		& 4300 		& 13000		& 130 		& 420 		& 1300 \\
  & &$\pm\frac13$		& 690 		& 2200 		& 6800 		& 64 		& 210 		& 670 \\
  & &$\pm\frac23$		& 350 		& 1100 		& 3400 		& 32 		& 110 		& 340 \\
  & &$\pm1$			& 230 		& 730 		& 2300 		& 22 		& 72 		& 230 \\
  & &$\pm2$			& 120 		& 370 		& 1200 		& 13 		& 43 		& 140 \\
  & &$\pm3$			& 80 		& 250 		& 790 		& - 		& - 		& - \\
 \hline
  &2&0				& 70 		& 230 		& 740 		& 11 		& 35 		& 110 \\
  & &$\pm\frac12$		& 69 		& 220 		& 720 		& 10 		& 34 		& 110 \\
  & &$\pm1$			& 65 		& 210 		& 670 		& 9.7 		& 32 		& 100 \\
  & &$\pm2$			& 54 		& 170 		& 550 		& - 		& - 		& - \\
 \hline
  &3&0				& 35 		& 110 		& 370 		& 6.0 		& 20 		& 64 \\
  & &$\pm1$			& 34 		& 110 		& 350 		& 6.1 		& 20 		& 65 \\
  & &$\pm2$			& 31 		& 100 		& 330 		& - 		& - 		& - \\
 \hline\hline
 3&1&0				& 54		& 190		& 700		& 17		& 56		& 190 \\
  & &$\pm\frac13$		& 54		& 200		& 710		& 17		& 60		& 210 \\
  & &$\pm\frac23$		& 56		& 210		& 750		& 21		& 77		& 300 \\
  & &$\pm1$			& 59		& 220		& 830		& 72		& 140		& 340 \\
  & &$\pm2$			& 110 		& 800		& 1600 		& - 		& - 		& - \\
 \hline
  &2&0				& 180 		& 350 		& 850 		& 13 		& 37 		& 110 \\
  & &$\pm1$			& 110 		& 250 		& 660 		& 9.0 		& 28 		& 84 \\
  & &$\pm2$			& 57 		& 150 		& 440 		& - 		& - 		& - \\
 \hline
  &3&0				& 29 		& 86 		& 260 		& - 		& - 		& - \\
  & &$\pm1$			& 27 		& 82 		& 250 		& - 		& - 		& - \\
  & &$\pm2$			& 24 		& 72 		& 220 		& - 		& - 		& - \\
 \hline\hline
 8&1&0				& 20		& 74		& 270		& 7.3		& 25		& 86 \\
  & &$\pm1$			& 21		& 77		& 280		& - 		& - 		& - \\
  & &$\pm2$			& 23		& 91		& 360		& - 		& - 		& - \\
 \hline
  &2&0				& 17 		& 67 		& 270 		& - 		& - 		& - \\
  & &$\pm1$			& 18 		& 72 		& 310 		& - 		& - 		& - \\
  & &$\pm2$			& 21 		& 110 		& 780 		& - 		& - 		& - \\
 \hline
  &3&0				& 63 		& 110		& 270 		& - 		& - 		& - \\
  & &$\pm1$			& 50 		& 98 		& 240 		& - 		& - 		& - \\
  & &$\pm2$			& 32 		& 72 		& 190 		& - 		& - 		& - \\
 \hline
 \end{tabular}
 \caption{Naturalness bounds on the mass $M$ (in TeV and to 2 significant figures)
 of various vector-like fermionic gauge multiplets
 for $\Lambda_h=\{M^+,10^{19}\text{ GeV}\}$ and $\Delta(\Lambda_h)=\{10,100,1000\}$.
 The dashes indicate that a Landau pole arises below $10^{19}$~GeV along the $\Delta(\Lambda_h)=10$ contour.
 }
\label{TabFermions}
 \end{minipage}
\end{table}

\begin{table}[p]
 \begin{minipage}[c][0.95\textheight]{\textwidth}
 \scriptsize
 \begin{tabular}{||ccc||c|c|c||c|c|c||}
 \hline
         &           &          & \multicolumn{3}{c||}{$\Lambda_h=M^+$}        & \multicolumn{3}{c||}{$\Lambda_h=10^{19}$~GeV} \\
 \hline
 $SU(3)$ & $SU(2)_L$ & $U(1)_Y$ & $\Delta=10$ 	& $\Delta=100$ 	& $\Delta=1000$ & $\Delta=10$ 	& $\Delta=100$ 	& $\Delta=1000$ \\
 \hline\hline
 1&1&$\pm\frac16$		&  1300		&  4100		&  13000	& 29 		& 96 		& 310 \\
  & &$\pm\frac13$		&  670		&  2000		&  6400		& 14 		& 47 		& 150 \\
  & &$\pm\frac23$		&  340		&  1000		&  3200		& 6.8 		& 23 		& 75 \\
  & &$\pm1$			&  230		&  690		&  2200		& 4.4 		& 15 		& 48 \\
  & &$\pm2$			&  120		&  350		&  1100		& 2.0 		& 6.5 		& 21 \\
  & &$\pm3$			&  77		&  240		&  740		& - 		& - 		& - \\
 \hline
  &2&0				& 67 		&  210		& 680 		& 2.3 		& 7.7 		& 25 \\
  & &$\pm\frac12$		& 65		&  210		& 660		& 2.1 		& 7.2		& 24 \\
  & &$\pm1$			& 62		&  190		& 620		& 1.8 		& 6.0		& 20 \\
  & &$\pm2$			& 52		&  160		& 510		& 1.1		& 3.6		& 12 \\
 \hline
  &3&0				& 33 		&  100		& 340 		& 1.1 		& 3.6 		& 12 \\
  & &$\pm1$			& 32 		&  100		& 330 		& 0.95 		& 3.2 		& 10 \\
  & &$\pm2$			& 30 		&  94		& 300 		& 0.45 		& 1.7 		& 6.7 \\
 \hline\hline
 3&1&0				& 220		& 820		& 2900		& 12		& 40		& 130 \\
  & &$\pm\frac13$		& 290		& 1200		& 5400		& 12		& 38		& 110 \\
  & &$\pm\frac23$		& 330		& 880		& 2500		& 4.2		& 14		& 45 \\
  & &$\pm1$			& 160		& 470		& 1400		& 2.5 		& 8.4		& 27 \\
  & &$\pm2$			& 71	 	& 210		& 660		& 0.99 		& 3.2		& 10 \\
 \hline
  &2&$\pm0$			&  40		&  130		&  400		&  1.3		& 4.3 		& 14  \\
  & &$\pm1$			&  37		&  120		&  370		&  0.99		& 3.3 		& 11 \\
  & &$\pm2$			&  31		&  96		&  300		& - 		& - 		& - \\
 \hline
  &3&$\pm0$			& 20		&  62		&  200		& - 		& - 		& - \\
  & &$\pm1$			& 19 		&  60		&  190		& - 		& - 		& - \\
  & &$\pm2$			& 18 		&  55		&  180		& - 		& - 		& - \\
 \hline
 \end{tabular}
 \caption[r]{Naturalness bounds on the mass $M$ (in TeV and to 2 significant figures)
 of various scalar gauge multiplets
 for $\Lambda_h=\{M^+,10^{19}\text{ GeV}\}$ and $\Delta(\Lambda_h)=\{10,100,1000\}$.
 The dashes indicate that a Landau pole arises below $10^{19}$~GeV along the $\Delta(\Lambda_h)=10$ contour.
 }
 \label{TabScalars}
 \end{minipage}
\end{table}

Before we discuss them in more detail, let us briefly reiterate their meaning.
The scale $\Lambda_h$ corresponds to the input scale of
$\overline{\rm MS}$ parameters in the SM+GM EFT.
The quantity $\Delta(\Lambda_h)$, defined in \eq{EqFTuning1}, is a sensitivity measure for the Higgs mass parameter
which can be interpreted as a Bayesian evidence on the Jeffreys scale
or (more loosely) to a percentage fine-tuning in the Barbieri--Giudice sense.
A stringent naturalness constraint is then $\Delta(\Lambda_{Pl})<10$,
which (loosely) ensures $<10\%$ sensitivity for $\mu^2(m_Z)\simeq -(88\text{ GeV})^2$
when the $\overline{\rm MS}$ parameters are defined at $\Lambda_{Pl}$.
If a phenomenological model can satisfy this constraint
then we would say it does not induce a Higgs naturalness problem.
The bounds weaken in the limit $\Lambda_h\to M^+$.
Still, rather remarkably, they remain finite in this limit,
as we argued in \sec{SecNaturalness}.
The $\Delta(M^+)<10$ bound can therefore be interpreted
as a conservative naturalness constraint on $M$.
It is also of interest if $\Delta(\Lambda_{Pl})$ is not applicable,
e.g. if new physics arises at a scale above $M$ 
which markedly affects the $\mu^2(\mu_R)$ evolution,
or if the EFT hits a Landau pole below $\Lambda_{Pl}$.

\section{Discussion \label{SecDiscussion}}

Some aspects of our results can be understood by scaling relations.
At fixed $\Lambda_h$, the bounds in \tabs{TabFermions} and \ref{TabScalars} 
scale approximately as $\sqrt{\Delta/C_2^{SM}}$,
as one would expect from \eq{EqDeltaToyModelF2} for the simple example
discussed in \sec{SecNaturalness}.
Where they are violated (particularly for the $\Delta(\Lambda_h)$ bounds)
it is due to some cancellation between contributions:
the contributions arising from $SU(3)$ charge are opposite in sign
to those from $SU(2)$ and $U(1)_Y$ charge.
The contour plots in \figs{FigFermionDelta}--\ref{FigScalarLambda} 
make these cancellations more obvious, 
and we will discuss them shortly.
Comparing bounds evaluated at disparate $\Lambda_h$ is more involved.
Indeed, this is why we have gone to the trouble of a two-loop RGE analysis!
Still, some qualitative observations will be made presently.

\begin{figure}[t]
  \centering
  \includegraphics[width=0.45\textwidth]{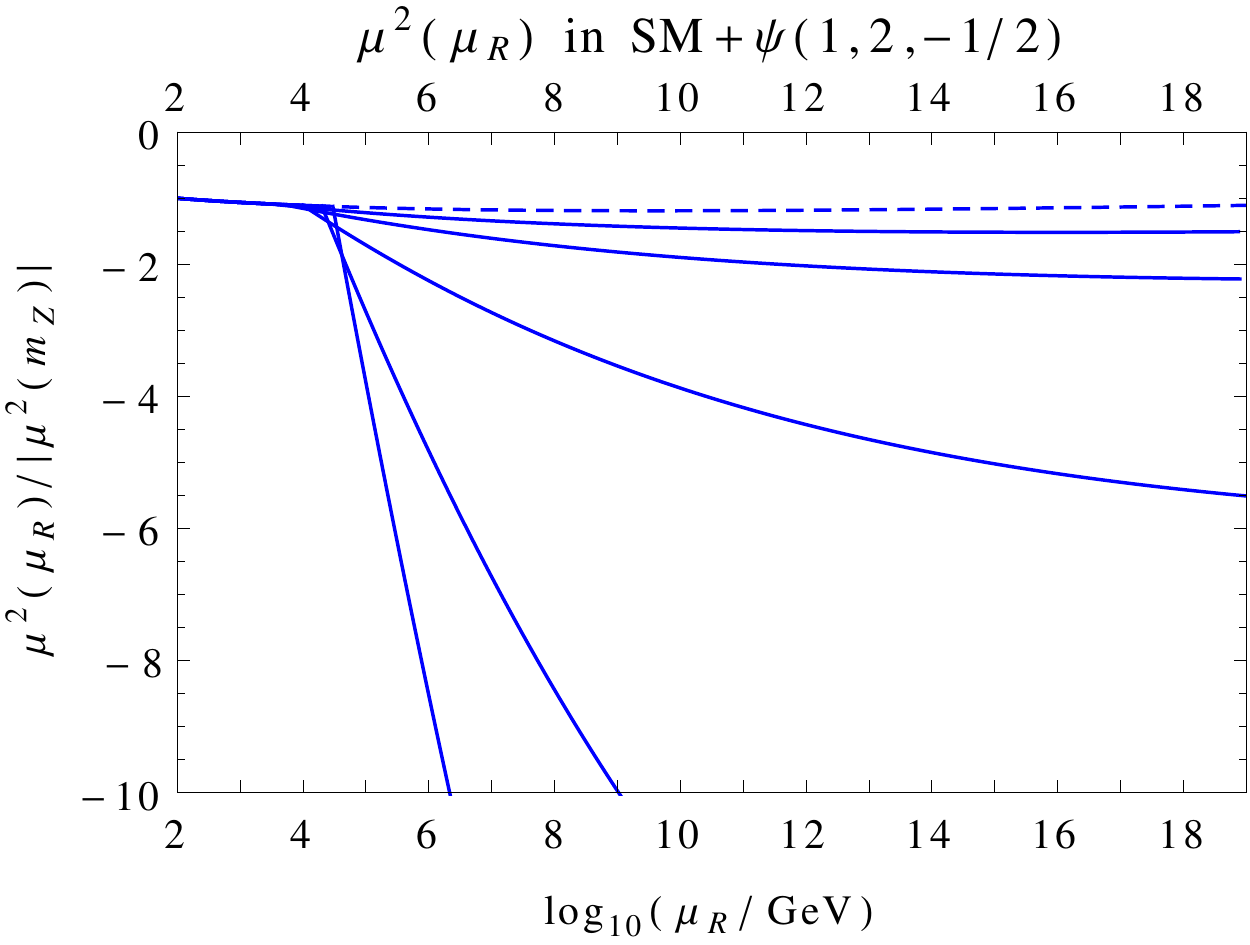}
  \includegraphics[width=0.45\textwidth]{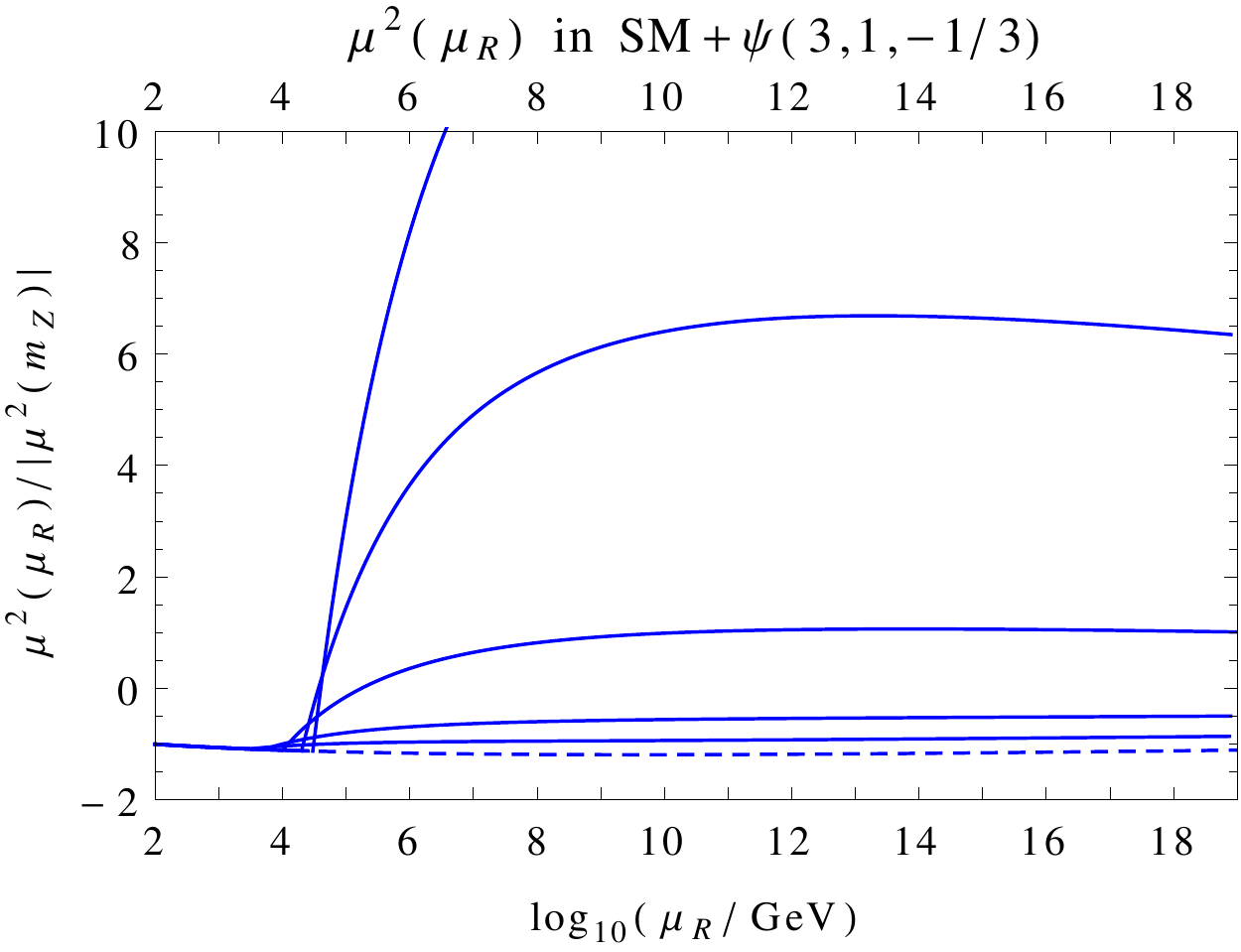}
  \caption{Example running of $\mu^2(\mu_R)$ in the SM+GM EFT for 
  a heavy lepton doublet $\psi(1,2,-1/2)$ and a heavy down-type quark $\psi(3,1,-1/3)$  with
  $M= 3,5,10,20,30$~TeV.
  The dashed line is the SM-only case.}
  \label{FigFermionRunning}
\end{figure}

For the fermionic GMs in \tab{TabFermions}, 
the rough scaling relation $\sim\sqrt{1/\sqrt{5}\log(\Lambda_{Pl}/M)}$
between bounds evaluated at $\Delta(M^+)$ and $\Delta(\Lambda_{Pl})$,
as expected from \eq{EqDeltaToyModelF2}, is broken by the RG evolution of $C_2$.
We observe that the naturalness bounds at $\Lambda_{Pl}$ are more stringent
than this relation would suggest for $\psi(1,1,Q_1)$,
and less stringent for $\psi(1,Q_2,0)$ and $\psi(Q_3,1,0)$.
This is simply because $g_1^4(\mu_R)$ (and therefore $C_2$) grows at higher energy,
whereas the opposite is true for $g_2^4(\mu_R)$ and $g_3^4(\mu_R) y_t^2(\mu_R)$.
This effect can be observed in \fig{FigFermionRunning}, 
where we show the example RG evolution of $\mu^2(\mu_R)$
for gauge multiplets of increasing mass.
The naturalness problem which broadly arises from 
a sensitivity to the high scale input 
$\mu^2(\Lambda_h)$ is self-evident for large masses.

\begin{figure}[t]
  \centering
   \includegraphics[width=0.46\textwidth]{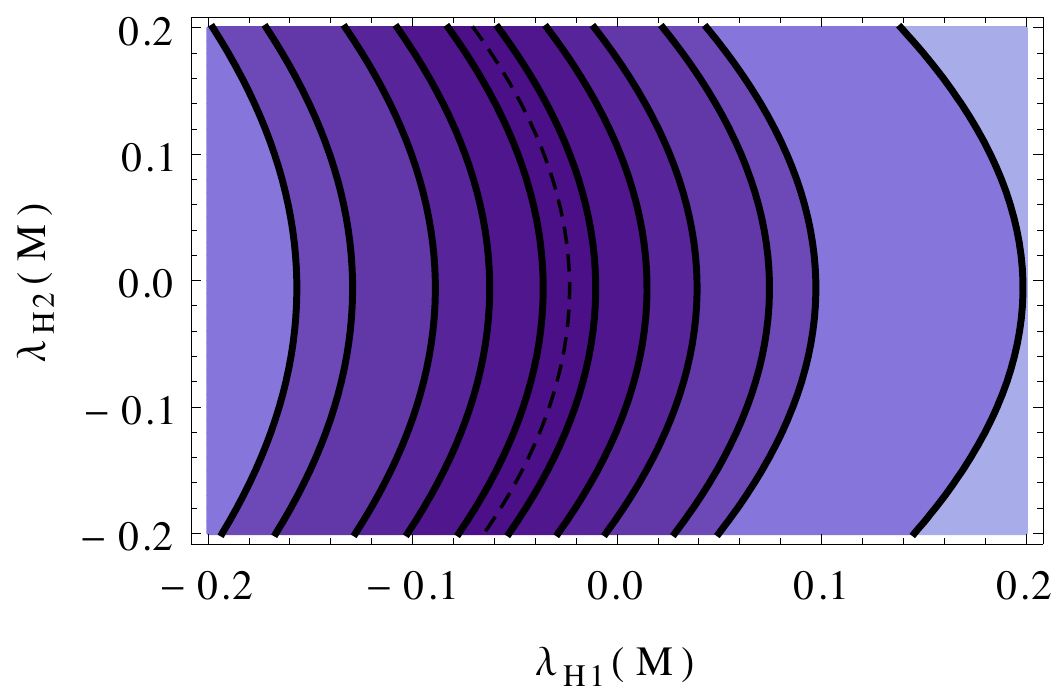}
   \includegraphics[width=0.46\textwidth]{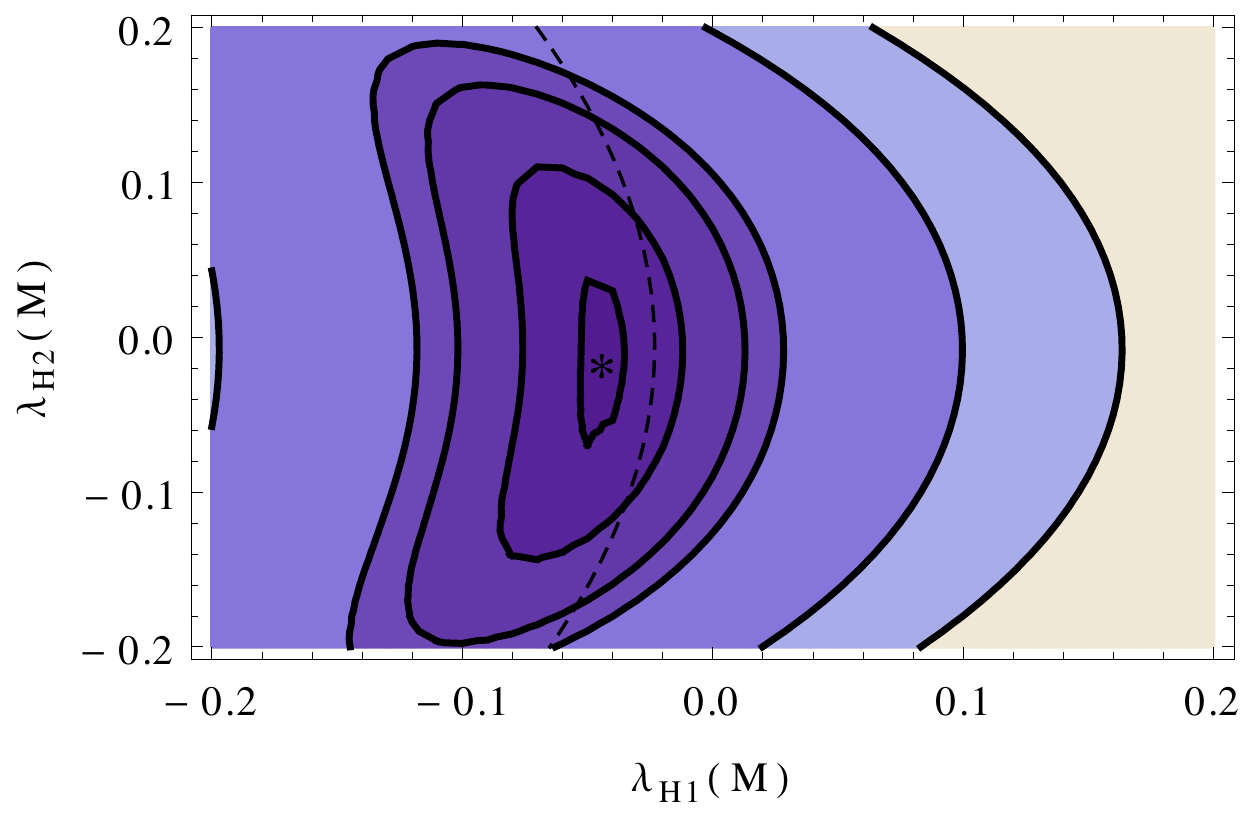}
   \put(10,30){\includegraphics[width=0.04\textwidth]{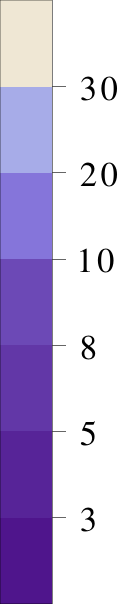}} \\
   \includegraphics[width=0.45\textwidth]{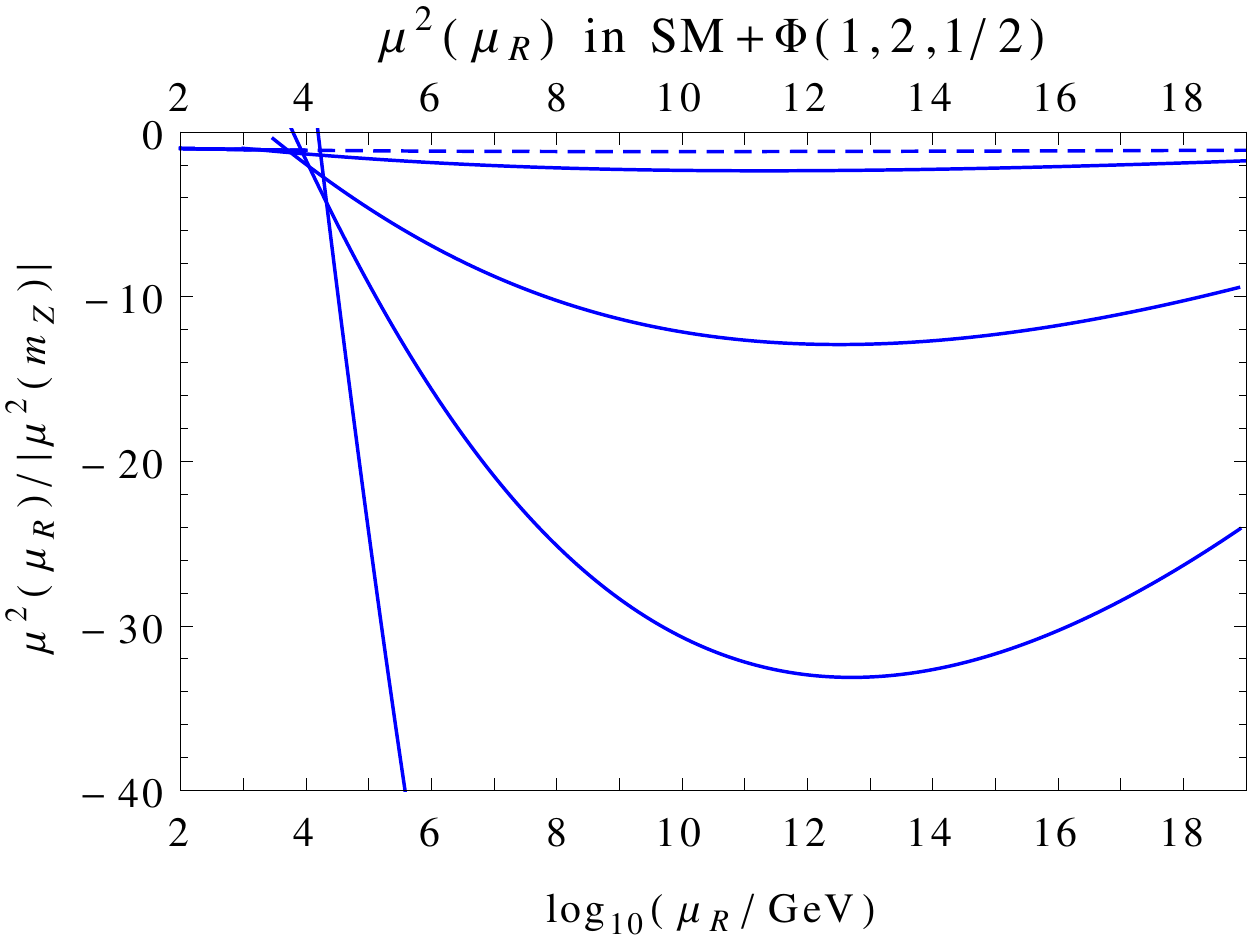}
   \includegraphics[width=0.45\textwidth]{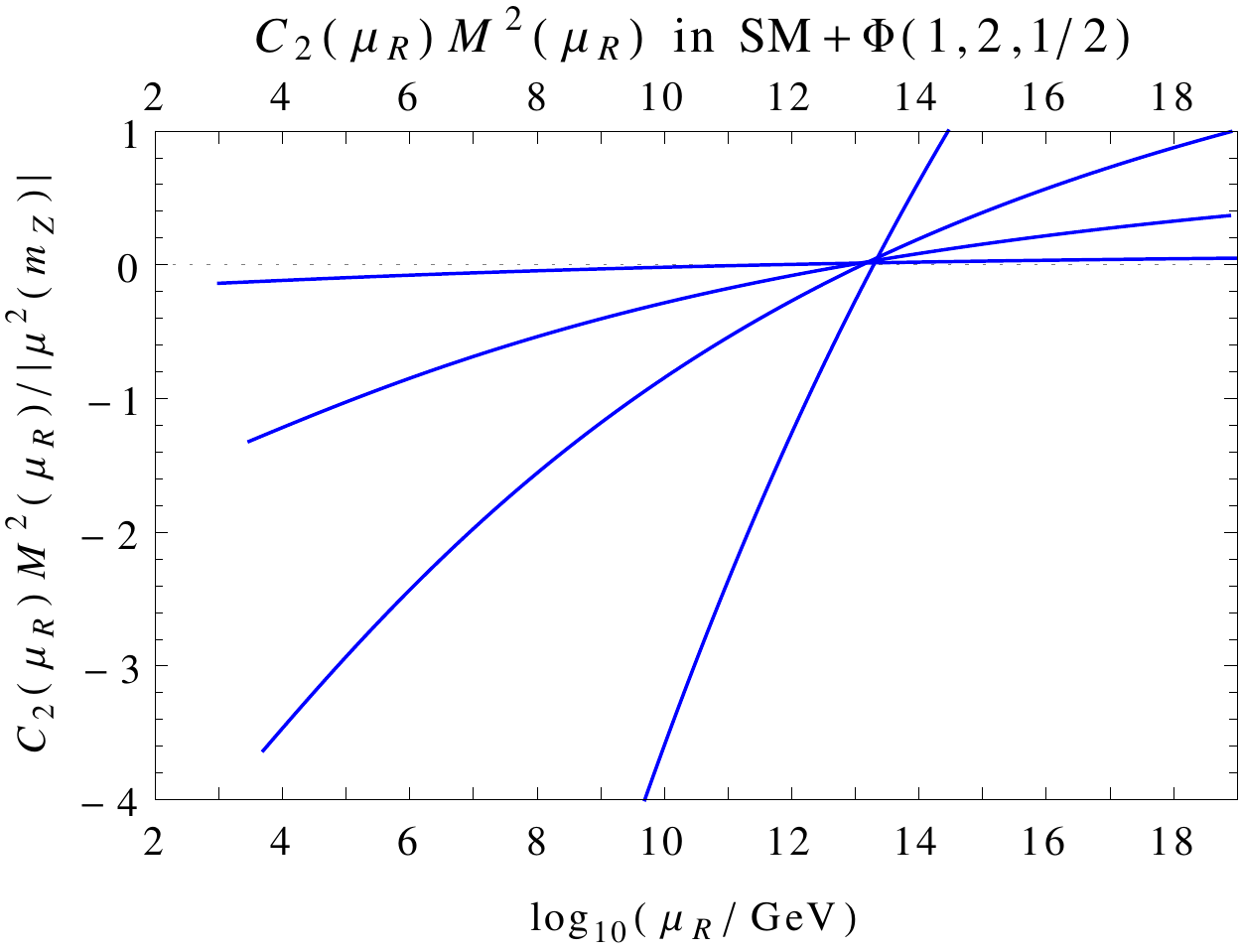}
  \caption{Sensitivity measure and RG evolution in the SM+$\Phi(1,2,1/2)$ (i.e. the 2HDM).
  Upper panel: The not-yet-minimised sensitivity measure 
    for $M=1$~TeV as a function of $(\lambda_{H1}(M),\lambda_{H2}(M))$ for:
   (left) $\Delta_{red}(\Lambda_{Pl})$ of \eq{EqSensReduced}, and 
   (right) $\Delta(\Lambda_{Pl})$ of \eq{EqFTuningScalar}.
    The dashed line shows the $\mu^2(\Lambda_{Pl})=0$ contour
    and the star denotes the global minimum.
  Lower panel: RG evolution of $\mu^2(\mu_R)$ and $C_2(\mu_R)M^2(\mu_R)$ for $M=1,3,5,10$~TeV
    evaluated at the $(\lambda_{H1},\lambda_{H2})$ 
    points which minimise $\Delta(\Lambda_{Pl})$.}
  \label{FigScalarExamples}
\end{figure}

For the scalar GMs in \tab{TabScalars}, 
there is no obvious scaling relation 
between naturalness bounds at different scales.
One observation is that,
although the $\Delta(M^+)$ bounds are similar\footnote{
The larger relative difference between the bounds for coloured states 
may be partly accounted for by the three-loop $g_3^4 y_t^2$ term
which is not captured in our pure two-loop scalar analysis.}
to those found in the fermionic GM case,
the $\Delta(\Lambda_{Pl})$ bounds are much more stringent.
This is because the sole one-loop term in the $\mu^2$ RGE
involves the portal quartic $\lambda_{H1}$, 
which is {\it itself} renormalised by pure gauge RG terms at one-loop.
Thus the scaling relation is expected to more closely resemble
$\sim 1/\log(\Lambda_{Pl}/M)$ [rather than $\sim \sqrt{1/\log(\Lambda_{Pl}/M)}$].
What actually happens is unfortunately quite opaque,
since it is hidden by various complexities: 
a coupled set of RGEs;
a non-trivial sensitivity measure \eq{EqFTuningScalar};
and a minimisation procedure over the $\lambda_{Hi}$.
Let us attempt to convey some intuition for what happens
by considering the example of a two Higgs doublet model, 
i.e. the SM+$\Phi(1,2,1/2)$.
To this end it is useful to define a reduced sensitivity measure
\begin{align}
 \Delta_{red}(M,\Lambda_h) = \left|
 \frac{\partial \log \mu^2(m_Z)}{\partial \log \mu^2(\Lambda_h)}
 \right| , \label{EqSensReduced}
\end{align}
which is a subcomponent of the full measure \eq{EqFTuningScalar}.
This reduced measure vanishes in the limit $\mu^2(\Lambda_h)\to 0$.
As we already argued in \sec{SecMethodScalar},
it is always possible to choose the $\lambda_{Hi}$ such that 
$C_2$ swaps sign over its RG evolution and $\mu^2(\Lambda_h)=0$.
Thus one expects a contour in $\lambda_{Hi}$ space 
along which $\Delta_{red}(M,\Lambda_h)$ vanishes \cite{Clarke2015hta}.
In \fig{FigScalarExamples} we plot $\Delta_{red}(M,\Lambda_h)$
as a function of $(\lambda_{H1}(M),\lambda_{H2}(M))$ 
for $\Lambda_h=\Lambda_{Pl}$ and $M=1$~TeV,
where such a contour is readily observed.
Obviously this contour constitutes a fine-tuning in the $\lambda_{Hi}$,
and we would hope that our full sensitivity measure captures this tuning
and restores a finite naturalness bound.
Indeed, it does; also shown in \fig{FigScalarExamples}
is the full sensitivity measure as a function of $(\lambda_{H1}(M),\lambda_{H2}(M))$,
with a unique minimum of $\Delta(\Lambda_{Pl})\simeq 2.7$
nearby the $\mu^2(\Lambda_h)=0$ contour.
In the lower panel of \fig{FigScalarExamples} we also show the running of 
$\mu^2(\mu_R)$ and $C_2(\mu_R)M^2(\mu_R)$ at this minimum
(and for other example masses).
It is seen that $C_2$ does switch sign, as expected.

We will now briefly comment on some features in the
$(M,\Delta)$ and $(M,\Lambda_h)$ contour plots of
\figs{FigFermionDelta}--\ref{FigScalarLambda}.
In \fig{FigFermionDelta} there is a sharp ``Veltman throat''
in the $\Delta(M^+)$ contours for coloured fermions.
This occurs when the three-loop colour contribution
cancels with the electroweak contributions such that $C_2(M)=0$.
It was already noted in \sec{Sec2Fermion}
that this is only an artifact of the loop level to which we are working.
The $\Delta(\Lambda_{Pl})$ contour for the $\psi(3,1,1)$ GM
demonstrates how this feature is effectively removed when $\Lambda_h >M$.
In \fig{FigScalarDelta} the qualitative form of the contours in the
$\Phi(3,1,Q_1)$ scalar case is seen to change as $Q_1$ is increased from 0 to 2.
This is due to a transition in dominance
between colour and hypercharge effects, which are opposite in sign.

In \fig{FigFermionLambda} a cusp feature is observed when $\Lambda_h$ is just above $M$.
This can be understood from the toy model \eq{EqDeltaToyModelF2}:
it is the point where $2\log(\Lambda_h/M)\simeq 1$ and the 
$\partial/\partial\log M^2$ sensitivity measure is minimised.
Also, the ``turn-around'' features in the $\psi(3,1,Q_1)$,
$\psi(8,1,Q_1)$, and $\psi(8,2,Q_1)$ plots
can again be understood as a balance between the colour and electroweak contributions.
In \fig{FigScalarLambda} a number of cusp features are observed,
mostly occuring at $\Lambda_h\sim 20 M$.
These features all have the same origin:
they occur for solutions where $\mu^2(\Lambda_h)\approx 0$. 
For example, in the $\Phi(1,2,1/2)$ case at $\Lambda_h=\Lambda_{Pl}$occurring
we saw that the $\lambda_{Hi}$ took on values such that $\mu^2(\Lambda_{Pl})<0$ (see \fig{FigScalarExamples}).
It turns out that, for $\Lambda_h\lesssim 20 M$, 
the sensitivity measure is minimised for values such that $\mu^2(\Lambda_h)>0$.
At the transition point the reduced sensitivity measure \eq{EqSensReduced} vanishes,
and hence the full sensitivity measure is somewhat reduced.
Note that in the cases where $\Phi$ is coloured the transition
occurs later due to the competing contributions between gauge contributions.

Before concluding we would like to make a few comments about the applicability 
of these bounds in the context of extended models. 
Firstly, one might contend that our bounds (especially the $\Delta(\Lambda_{Pl})$ bounds),
which are only derived in the context of minimal SM+GM extensions,
are not applicable in a realistic model with additional high scale states.
This is true in a quantitative sense: the bounds are sure to change.
Nonetheless, this does not imply that they are not qualitatively useful.
It would take very special physics 
to ameliorate these bounds by a significant amount.
For example, one could try to introduce new states with particular properties at $\Lambda_h\sim M$
such that loop contributions approximately cancel at this scale \cite{Fabbrichesi:2015zna}.
In the absence of a symmetry which introduces the appropriate correlations between parameters at this scale,
and a symmetry which ensures the cancellation remains satisfied under RG evolution,
naturalness bounds similar to those we have derived will be quickly reintroduced at $\Lambda_h > M$.
Actually, such symmetry requirements 
are just those provided by supersymmetric theories,
and herein lies the connection between our RG description 
and the usual naturalness arguments in the context of supersymmetry.
In any case, the framework we have outlined in Appendix~\ref{AppBayes}
is fully generalisable to perturbative models with more states.
Naturalness of the low scale Higgs mass parameter 
can be quantified by the Bayesian sensitivity measure \eq{EqBayesFac},
as long as one is prepared to calculate and solve
RGEs at least at two-loop order 
with one-loop matching between intermediate physical scales.

Secondly, in deriving our sensitivity measure we have made the assumption of flat priors
on the logarithms of $\overline{\rm MS}$ input parameters at scale $\Lambda_h$.
This particular set of priors is determined by
demanding insensitivity to units or parameter rescalings,
and makes logical sense in a bottom-up approach where one would like to remain 
maximally agnostic to the higher scale UV theory. 
However, it is true that 
if one were to derive these priors as posteriors arising from
a flat set of priors in the UV theory,
they would almost certainly not be flat. Furthermore,
one would generally expect correlations between the parameters.
Hence one might expect that our results 
are only broadly applicable if those derived priors are approximately flat.\footnote{%
In particular, some might argue that this is unlikely for $\log\mu^2(\Lambda_h)$ in the presence of gravity,
but then some might choose to remain agnostic.}

Nevertheless, it is still possible to argue that the naturalness bounds
derived here are not expected to significantly change even if the priors are peaked.
Consider for example altering the $\log\mu^2(\Lambda_h)$ prior such that it is
locally scaled by a factor $\kappa$ within some window;
i.e. the prior is instead flat in a function $f(\log\mu^2)$ with
$\partial f(\log\mu^2)/\partial\log\mu^2=\kappa$ (1) within (outside of) the window.
In such a case the sensitivity measure, e.g. \eq{EqFTuningScalar},
is of the same form except with $\partial\log\mu^2(\Lambda_h)$ 
replaced by $\partial f(\log\mu^2(\Lambda_h))$.
Now, consider the case where the prior is locally increased ($\kappa>1$) 
within a window centered on the realised value of $\log\mu^2(\Lambda_h)$. 
Then the contribution of the $|\partial\log\mu^2(m_Z)/\partial\log\mu^2(\Lambda_h)|$ term 
to the sensitivity measure will be scaled down by a factor $1/\kappa$.
However, the contribution from the other terms does not change.
Generally, as can be understood from our toy examples (see \eq{EqDeltaToyModelF2}),
the contribution from other terms 
(and at very least the $|\partial\log\mu^2(m_Z)/\partial\log M|$ term)
is of similar order.
Thus, in this case, one does not expect the bound to significantly change
unless the the prior in $\log M$ (and other parameters) is also locally peaked.
This is not surprising, since we have increased the probability of a specific
initial high scale boundary value for $\mu^2(\Lambda_h)$,
but we have not altered the prior on the slope,
which is controlled by the size of $M$ and other parameters entering the RGEs.
Thus if the bound is to be significantly affected by a set of peaked priors at high scale,
they need to be sharply peaked at very particular values in more than one parameter.
Unless one has a plausible explanation for such priors
then this has only shifted the naturalness issue.

\section{Conclusion \label{SecConclusion}}

The aim of this paper was to confront the question,
{\it at what mass does a heavy gauge multiplet introduce a 
physical Higgs naturalness problem?}
In \sec{SecNaturalness} we described a physical way 
to understand the Higgs naturalness problem
which might be introduced when perturbative 
heavy new physics is added to the SM. 
The description is of particular interest in bottom-up extensions of the SM.
The premise is essentially as follows.
In any perturbative EFT,
the low scale Higgs mass parameter $\mu^2(m_Z)\simeq -(88\text{ GeV})^2$
can be connected by renormalisation group equations to
$\overline{\rm MS}$ ``input'' parameters defined at some high scale $\Lambda_h$.
If $\mu^2(m_Z)$ is especially sensitive to these input parameters,
then this signifies a Higgs naturalness problem.
In particular, this can occur if a heavy particle of mass $M$ is added to the SM.

In order to sensibly quantify this potential problem,
we derived a sensitivity measure using Bayesian probabilistic arguments.
The measure can be interpreted as a Bayesian model comparison
[see \eq{EqFTuning0}]
which captures the ``naturalness price'' 
paid for promoting the Higgs mass parameter
to a high scale input parameter of the model
as opposed to a purely phenomenological input parameter at low scale.
It is fully generalisable to any perturbative QFT,
with the details provided in Appendix~\ref{AppBayes}.
The measure reduces in a certain (relevant) limit
to an intuitively motivated Barbieri--Giudice-like fine-tuning measure
[see \eq{EqFTuning2}].
The resulting sensitivity measure 
is generally a function of unknown high scale inputs.
We conservatively projected these out by minimising over them, 
thereby obtaining the sensitivity measure \eq{EqFTuning1},
which is a function of $\Lambda_h$ 
and the mass $M$ of a heavy new particle.

This sensitivity measure was used to set naturalness bounds
on the masses of various gauge multiplets,
using a full two-loop RGE analysis with one-loop matching.
An interesting outcome is that,
once RG effects are taken into account
and finite threshold corrections are captured,
a naturalness bound on $M$ remains
even in the limit $\Lambda_h\to M^+$.
The resulting bounds are presented in \tabs{TabFermions} and~\ref{TabScalars},
and as contours in \figs{FigFermionDelta}--\ref{FigScalarLambda}.
They form the main result of this paper,
and we hope they are of interest to model builders.
For $\Lambda_h=\Lambda_{Pl}$ we find ``$10\%$ fine-tuning'' bounds 
of $M<\mathcal{O}(1$--$10)$~TeV on the masses of 
various gauge multiplets,
with the bounds on fermionic gauge multiplets significantly weaker than for scalars.
In the limit $\Lambda_h\to M^+$ the bounds weaken to $M<\mathcal{O}(10$--$100)$~TeV;
these can be considered as conservative naturalness bounds,
of interest if new physics is expected
to substantially alter the RG evolution of $\mu^2(\mu_R)$
above the scale $M$.
We also found that the bounds on coloured multiplets are no more severe
than on electroweak multiplets, 
since they correct the Higgs mass directly at three-loop order.

\acknowledgments

This work was supported in part by the Australian Research Council.
JDC thanks the Max-Planck-Institut f{\"u}r Kernphysik in Heidelberg for their hospitality 
during the completion of this manuscript.


\bibliography{references}

\appendix

\newpage

\section{Sensitivity measure as a Bayesian model comparison \label{AppBayes}}

In this Appendix we show how a Barbieri--Giudice-like fine-tuning measure for $\mu^2(m_Z)$
arises in a certain limit of our Bayesian model comparison.
Similar connections have been made in earlier works,
e.g. Refs.~\cite{Cabrera:2008tj,Fichet2012sn}.

Bayesian probability allows one to assign 
a degree of belief to some hypothesis, 
in our case a particle physics model.
The model $\mathcal{M}$ consists of a set of input parameters $\mathcal{I}$
and a rule for connecting these to a set of observables $\mathcal{O}$.
Let us assume that there are $n$ fundamental input parameters
$\mathcal{I}=\{\mathcal{I}_1,\dots,\mathcal{I}_n\}$
and $m\le n$ independent observables 
$\mathcal{O}=\{\mathcal{O}_1,\dots,\mathcal{O}_m\}$.
The rule is just a map 
$\mathcal{R}:\mathcal{I}\to \mathcal{O}$
from input space to observable space with
$(\mathcal{I}_1,\dots,\mathcal{I}_n)\mapsto 
\mathcal{R}(\mathcal{I}_1,\dots,\mathcal{I}_n) = (\mathcal{O}_1,\dots,\mathcal{O}_m)$.
In this paper the models consist of the SM plus a new gauge multiplet of mass $M$,
with inputs as the logarithms of $\overline{{\rm MS}}$ parameters of the full Lagrangian defined at scale $\Lambda_h$,
observables as the logarithms of $\overline{{\rm MS}}$ SM Lagrangian parameters at scale $m_Z$,
and $\mathcal{R}$ given by the RGEs.
The logarithms are taken to avoid dependence on units or rescalings of the Lagrangian.\footnote{%
Absolute values inside the logarithms are implied.
The signs of the parameters can be considered as separate inputs.
Explicitly including them with a flat prior probability mass function
does not change the final result, and we ignore them henceforth for clarity.}

The Bayesian evidence $B$ for $\mathcal{M}$
is the probability that the observables $\mathcal{O}$ 
attain their experimentally observed values $\mathcal{O}_{ex}$,
assuming $\mathcal{M}$ is true:
\begin{align}
 B(\mathcal{M}) := p(\mathcal{O}=\mathcal{O}_{ex} | \mathcal{M}) =
  \int p( \mathcal{O}=\mathcal{O}_{ex} | \mathcal{I} )
   \ p( \mathcal{I} ) 
   \ d\mathcal{I} 
   \ , \label{EqBayesEv}
\end{align}
where $p( \mathcal{O}=\mathcal{O}_{ex} | \mathcal{I} )$
is also called the likelihood function
$\mathcal{L}(\mathcal{I})$,
and $p(\mathcal{I})$ is the prior density for the model parameters.
The prior density represents the degree of belief 
in the values of the input parameters before any observations are made.
In the absence of any knowledge about the complete UV theory,
we should assume priors which are maximally agnostic.
This corresponds to a flat prior in the 
$n$-dimensional input space $\mathcal{I}$.
The mapping $\mathcal{R}$ can be used to express some point in input space
$(\mathcal{I}_1,\dots,\mathcal{I}_n)$
in terms of a new set of coordinates
$(\mathcal{O},\mathcal{I}')\equiv (\mathcal{O}_1,\dots,\mathcal{O}_m,\mathcal{I}_{m+1},\dots,\mathcal{I}_n)$
simply by
$\mathcal{I}
\mapsto 
\mathcal{R'}(\mathcal{I})
\equiv(\mathcal{R}(\mathcal{I}),\mathcal{I}')$.
We assume that this is a one-to-one mapping
(indeed, it is for RGEs in the perturbative regime).
If we assume perfectly measured observables,
then \eq{EqBayesEv} becomes
\begin{align}
B(\mathcal{M}) \propto
  \int \delta( \mathcal{O}-\mathcal{O}_{ex} )
   \ p\circ\mathcal{R}'^{-1}(\mathcal{O},\mathcal{I}')
 \left|\left(
 \begin{array}{ccc}
  \frac{\partial\mathcal{O}_1}{\partial\mathcal{I}_1} & \cdots & \frac{\partial\mathcal{O}_1}{\partial\mathcal{I}_m} \\
  \vdots & \ddots & \vdots \\
  \frac{\partial\mathcal{O}_m}{\partial\mathcal{I}_1} & \cdots & \frac{\partial\mathcal{O}_m}{\partial\mathcal{I}_m}
 \end{array}
 \right)\right|^{-1}
 d\mathcal{O}_1\cdots d\mathcal{O}_m d\mathcal{I}_{m+1} \cdots d\mathcal{I}_n
   \ ,
\end{align}
where the likelihood has become a delta function multiplied by a constant term,
and $|\left(\ \cdot \ \right)|\equiv |{\rm det}[(\ \cdot\ )] |$ is the determinant of the Jacobian
associated with the coordinate transformation.
Performing the integration over the observables,
\begin{align}
 \left.
 B(\mathcal{M})\propto
  \int p'(\mathcal{I}')
 \left|\left(
 \begin{array}{ccc}
  \frac{\partial\mathcal{O}_1}{\partial\mathcal{I}_1} & \cdots & \frac{\partial\mathcal{O}_1}{\partial\mathcal{I}_m} \\
  \vdots & \ddots & \vdots \\
  \frac{\partial\mathcal{O}_m}{\partial\mathcal{I}_1} & \cdots & \frac{\partial\mathcal{O}_m}{\partial\mathcal{I}_m}
 \end{array}
 \right)\right|^{-1}
   d\mathcal{I}_{m+1} \cdots d\mathcal{I}_n 
   \ \right|_{\mathcal{O}=\mathcal{O}_{ex}}
   \ ,
\end{align}
where $p'(\mathcal{I}')\equiv p\circ\mathcal{R}^{-1}(\mathcal{O}_{ex},\mathcal{I}')$.
The requirement $\mathcal{O}=\mathcal{O}_{ex}$
has carved out an experimentally allowed $(n-m)$ dimensional submanifold within the original $n$ dimensional input space.
We know that, since the original prior was flat in $n$ dimensions,
the prior on the submanifold must be flat with respect to the induced volume element
(as opposed to the volume element $d\mathcal{I}_{m+1} \cdots d\mathcal{I}_n$).
We can rescale the existing volume element to write, equivalently,
\begin{align}
\left.
 B(\mathcal{M}) \propto
  \int p'(\mathcal{I}')
 \frac{
 \left|\left(
 \begin{array}{ccc}
  \frac{\partial\mathcal{O}_1}{\partial\mathcal{I}_1} & \cdots & \frac{\partial\mathcal{O}_1}{\partial\mathcal{I}_m} \\
  \vdots & \ddots & \vdots \\
  \frac{\partial\mathcal{O}_m}{\partial\mathcal{I}_1} & \cdots & \frac{\partial\mathcal{O}_m}{\partial\mathcal{I}_m}
 \end{array}
 \right)\right|^{-1}}
 {\sqrt{\left|
 \left(
 \begin{array}{ccc}
  \frac{\partial\mathcal{I}_1}{\partial \mathcal{I}_{m+1}} & \cdots & \frac{\partial\mathcal{I}_1}{\partial \mathcal{I}_n} \\
  \vdots & \ddots & \vdots \\
  \frac{\partial\mathcal{I}_n}{\partial \mathcal{I}_{m+1}} & \cdots & \frac{\partial\mathcal{I}_n}{\partial \mathcal{I}_n}
 \end{array}
 \right)^T
 \left(
 \begin{array}{ccc}
  \frac{\partial\mathcal{I}_1}{\partial \mathcal{I}_{m+1}} & \cdots & \frac{\partial\mathcal{I}_1}{\partial \mathcal{I}_n} \\
  \vdots & \ddots & \vdots \\
  \frac{\partial\mathcal{I}_n}{\partial \mathcal{I}_{m+1}} & \cdots & \frac{\partial\mathcal{I}_n}{\partial \mathcal{I}_n}
 \end{array}
 \right) 
 \right|}}
   \ d\Sigma 
   \ \right|_{\mathcal{O}=\mathcal{O}_{ex}}
   \ ,
\end{align}
where the quantity under the square root is the determinant of the induced metric,
$d\Sigma$ is the induced volume element, 
and the prior $p'(\mathcal{I}')$ is constant with respect to this volume element.
This reduces to\footnote{%
To show this requires the use of:
Sylvester's identity
${\rm det}(\mathbb{I}_m+AB)={\rm det}(\mathbb{I}_n+BA)$
for $m\times n$ and $n\times m$ matrices $A$ and $B$;
and the matrix identity
$(A|B)(A|B)^T=AA^T+BB^T$
for $A$ and $B$ matrices with equal number of rows.}
\begin{align}
 \left.
 B(\mathcal{M}) \propto
  \int \frac{p'(\mathcal{I}')}{\sqrt{\left| J J^T \right|}} \  d\Sigma 
  \ \right|_{\mathcal{O}=\mathcal{O}_{ex}}  \ ,
  \label{EqBayesWithJTJNo1}
\end{align}
where $J$ is the $m\times n$ matrix 
defined by $J_{ij} = \partial\mathcal{O}_i / \partial\mathcal{I}_j$ \cite{Fichet2012sn}.
Additionally, by taking a delta function prior on $\mathcal{I}'$
we can evaluate (and compare) Bayesian evidence for the model $\mathcal{M}$ 
with unconstrained input parameters 
$(\mathcal{I}_{m+1},\dots,\mathcal{I}_n)$ taking on specific values:
\begin{align}
 \left.
 B(\mathcal{M};\mathcal{I}') \propto
  \frac{1}{\sqrt{\left| J J^T \right|}}
  \ \right|_{
  ^{\mathcal{O}_{ex}}
  _{\mathcal{I}'}
  }  \ .
  \label{EqBayesWithJTJ}
\end{align}

Let us now put this in the context of minimal extensions of the SM by a gauge multiplet of mass $M$.
We take $\mathcal{I}_1=\log\mu^2(\Lambda_h)$ and $\mathcal{O}_1=\log\mu^2(m_Z)$.
The remaining inputs and observables are logarithms of the $\overline{\rm MS}$ Lagrangian parameters.
The Bayesian evidence \eq{EqBayesWithJTJ} is not enough by itself;
it can only be interpreted with respect to some reference model.
We will, after all, be interested in the sensitivity of $\mu^2(m_Z)$ to the input parameters,
and we have not so far treated the $\mu^2$ parameter in any special way.
The reference model we choose to compare to is the model $\mathcal{M}_0$ 
in which the Higgs mass parameter is instead taken as a ``phenomenological'' input parameter at scale $m_Z$,
i.e. $\mathcal{I}_1=\mathcal{O}_1=\log\mu^2(m_Z)$.
In $\mathcal{M}_0$ we have that $J_{11}=1$ and $J_{1j}=0$ for $j>1$.
The Bayes factor between these two models is then
\begin{align}
 \left.
 K(\mathcal{M};\mathcal{I}') := \frac{B(\mathcal{M}_0;\mathcal{I}')}{B(\mathcal{M};\mathcal{I}')}
  \right|_{
  ^{\mathcal{O}_{ex}}
  _{\mathcal{I}'}
  } . \label{EqBayesFac}
\end{align}  
Since the dimensionful parameter $\mu^2(\mu_R)$ does not enter 
the (mass independent) RGEs of the remaining dimensionless observables,
we have that $J_{i1}=\partial\mathcal{O}_i / \partial\mathcal{I}_1=0$ for $i>1$.
Additionally, in the special case 
that the dimensionless observables
are approximately insensitive to the unconstrained inputs,
i.e. $J_{ij}\simeq 0$ for $i>1$ and $j\ge m+1$,
\eq{EqBayesWithJTJ} becomes
\begin{align}
 B(\mathcal{M};\mathcal{I}') \propto
 \left.
 \frac{
   \left|\left(
 \begin{array}{ccc}
  \frac{\partial\mathcal{O}_2}{\partial\mathcal{I}_2} & \cdots & \frac{\partial\mathcal{O}_2}{\partial\mathcal{I}_m} \\
  \vdots & \ddots & \vdots \\
  \frac{\partial\mathcal{O}_m}{\partial\mathcal{I}_2} & \cdots & \frac{\partial\mathcal{O}_m}{\partial\mathcal{I}_m}
 \end{array}
 \right)\right|^{-1}
  }
  {\sqrt{
   \left( \frac{\partial\log\mu^2(m_Z)}{\partial\mathcal{I}_1} \right)^2 
   + \sum\limits_{j\ge m+1} \left( \frac{\partial\log\mu^2(m_Z)}{\partial\mathcal{I}_j} \right)^2
   } 
   }
  \ \right|_{
  ^{\mathcal{O}_{ex}}
  _{\mathcal{I}'}
  }  \ .
\end{align}
We can see that a Barbieri--Giudice-like fine-tuning measure has appeared in the denominator.
In this case the quantity $B(\mathcal{M}_0;\mathcal{I}')$ becomes independent of $\mathcal{I}'$,
and the Bayes factor \eq{EqBayesFac} is
\begin{align}
 \left.
 K(\mathcal{M};\mathcal{I}') 
 = \sqrt{
   \left( \frac{\partial\log\mu^2(m_Z)}{\partial\log\mu^2(\Lambda_h)} \right)^2
   + \sum\limits_{j\ge m+1} \left( \frac{\partial\log\mu^2(m_Z)}{\partial\mathcal{I}_j} \right)^2
   }
  \right|_{
  ^{\mathcal{O}_{ex}}
  _{\mathcal{I}'}
  }.
\end{align}
This is reminiscent of the Barbieri--Giudice fine-tuning measure.
We observe the interesting emergence of 
additional terms quantifying Higgs mass sensitivity 
only to the unconstrained parameters of the model.
Conceptually, $K$ is a comparison between
a flat prior in $\log\mu^2(m_Z)$ and
the RG devolved (to $m_Z$) flat prior in $\log\mu^2(\Lambda_h)$,
in the vicinity $\mu^2(m_Z)\simeq -(88\text{ GeV})^2$.
A Bayes factor of $K>10$ corresponds to the onset of strong evidence (on the Jeffreys scale)
for $\mathcal{M}_0$ over $\mathcal{M}$.

Lastly, note that the Bayes factor $K$ is still a function of the unknown parameters $\mathcal{I}'$.
In order to write down a sensitivity measure for the model $\mathcal{M}$
as a function of a subset of these unknown parameters (e.g. the gauge multiplet mass $M$),
we might want some way of projecting out the nuisance unknowns.
One way is to integrate over some region of $\mathcal{I}'$,
i.e. evaluate \eq{EqBayesWithJTJNo1}.
However, in this paper we instead choose the following conservative projection:
\begin{align}
 \Delta(\mathcal{M}) = \min\limits_{\mathcal{I}'} \bigg\{ K(\mathcal{M};\mathcal{I}') \bigg\} .
\end{align}
This identifies the best case scenario for Higgs mass naturalness in $\mathcal{M}$
by finding the point in $\mathcal{I}'$ with the lowest Bayes factor.

\end{document}

%% file: tikzdefs.tex
\usepackage{tikz}

\usetikzlibrary{external}
\tikzsetexternalprefix{tikzfigs/}
\usetikzlibrary{arrows,calc,graphs,patterns,positioning}
\usetikzlibrary{decorations,decorations.markings,decorations.pathmorphing,decorations.pathreplacing}
\usetikzlibrary{shapes.geometric}

\tikzset{/pgf/decoration/.cd,
    number of sines/.initial=5,
    angle step/.initial=20,
}
\newdimen\tmpdimen

\pgfdeclaredecoration{complete sines}{initial}
{
    \state{initial}[
        width=+0pt,
        next state=move,
        persistent precomputation={
            \pgfmathparse{\pgfkeysvalueof{/pgf/decoration/angle step}}%
            \let\anglestep=\pgfmathresult%
            \let\currentangle=\pgfmathresult%
            \pgfmathsetlengthmacro{\pointsperanglestep}%
                {(\pgfdecoratedremainingdistance/\pgfkeysvalueof{/pgf/decoration/number of sines})/360*\anglestep}%
        }] {}
    \state{move}[width=+\pointsperanglestep, next state=draw]{
        \pgfpathmoveto{\pgfpointorigin}
    }
    \state{draw}[width=+\pointsperanglestep, switch if less than=1.25*\pointsperanglestep to final, 
        persistent postcomputation={
        \pgfmathparse{mod(\currentangle+\anglestep, 360)}%
        \let\currentangle=\pgfmathresult%
    }]{%
        \pgfmathsin{+\currentangle}%
        \tmpdimen=\pgfdecorationsegmentamplitude%
        \tmpdimen=\pgfmathresult\tmpdimen%
        \divide\tmpdimen by2\relax%
        \pgfpathlineto{\pgfqpoint{0pt}{\tmpdimen}}%
    }
    \state{final}{
        \ifdim\pgfdecoratedremainingdistance>0pt\relax
            \pgfpathlineto{\pgfpointdecoratedpathlast}
        \fi
   }
}

\tikzset{
    scalar/.style={draw=black, thick, dashed},
    fermion/.style={draw=black, postaction={decorate}, decoration={markings, mark=at position .65 with {\arrow[]{latex}}}, thick},
    antifermion/.style={draw=black, postaction={decorate}, decoration={markings, mark=at position .65 with {\arrow{latex}}}, thick},
    photon/.style={draw=none, outer ysep=3pt, postaction={draw=black, thick, decorate}, decoration={complete sines, amplitude=5pt}},
    photonloop/.style={draw=none, outer ysep=3pt, postaction={draw=black, thick, decorate}, decoration={complete sines, number of sines=15, amplitude=5pt}},
    gluon/.style={draw=none, outer ysep=3pt, postaction={draw=black, thick, decorate}, decoration={coil, amplitude=3pt, segment length=5pt}}, 
    composite/.style={draw=gray!70!, line width=5pt},
    vtx/.style={inner sep=0pt},
    ins/.style={draw, cross, shape=circle, minimum size=5pt, inner sep=0pt}, 
    cross/.style={path picture={\draw[black] (path picture bounding box.south east) -- (path picture bounding box.north west) (path picture bounding box.south west) -- (path picture bounding box.north east);}},
    vertex/.style={draw, shape=circle, fill=black, minimum size=3pt, inner sep=0pt},
    blob/.style={draw, shape=circle, preaction={fill,white}, pattern = north west lines, minimum size=15pt, inner sep=0pt},        
    loop/.style={shape=circle, minimum size=1.7cm, outer sep=9pt},
    site/.style={draw, shape=circle, minimum size=1.5cm, inner sep=0pt},
    brane/.style={trapezium, draw, trapezium stretches=true, minimum height=4cm, minimum width=3.5cm, inner sep=0, trapezium left angle=35, trapezium right angle=145, shape border uses incircle, shape border rotate=17.5},
    momentum/.style={->, dist/.store in=\segDistance, pos/.store in=\segPos, len/.store in=\segLength,
      to path={
      ($(\tikztostart)!\segPos!(\tikztotarget)!\segLength/2!(\tikztostart)!\segDistance!90:(\tikztotarget)$) -- 
      ($(\tikztostart)!\segPos!(\tikztotarget)!\segLength/2!(\tikztotarget)!\segDistance!-90:(\tikztostart)$)  \tikztonodes
      }, 
      pos=.5,
      len=7mm,
      dist=2mm
    },    
}